\documentclass[preprint]{aastex}
\received{}
\revised{}
\accepted{}
\journalid{}{}
\articleid{}{}
\paperid{}
\cpright{}{}
\ccc{}
\lefthead{Matthews et al.}
\righthead{Detections of CO in LSB Spirals}
\begin{document}

\newcommand{\msun}{\mbox{${\cal M}_\odot$}}
\newcommand{\lsun}{\mbox{${\cal L}_\odot$}}
\newcommand{\kms}{\mbox{km s$^{-1}$}}
\newcommand{\HI}{\mbox{H\,{\sc i}}}
\newcommand{\HA}{H$\alpha$}
\newcommand{\mhi}{\mbox{$M_{\rm HI}$}}
\newcommand{\HII}{\mbox{H\,{\sc ii}}}
\newcommand{\am}[2]{$#1'\,\hspace{-1.7mm}.\hspace{.0mm}#2$}
\newcommand{\as}[2]{$#1''\,\hspace{-1.7mm}.\hspace{.0mm}#2$}
\newcommand{\ad}[2]{$#1^{\circ}\,\hspace{-1.7mm}.\hspace{.0mm}#2$}
\newcommand{\lsim}{~\rlap{$<$}{\lower 1.0ex\hbox{$\sim$}}}
\newcommand{\gsim}{~\rlap{$>$}{\lower 1.0ex\hbox{$\sim$}}}
\newcommand{\dark}{$M_{HI}/L_{B}$}
\newcommand{\tm}{$\times$}

   \title{Detections of CO in Late-Type, Low Surface Brightness Spiral Galaxies}

  \author{Lynn D. Matthews\altaffilmark{1}}
           \author{Yu Gao\altaffilmark{2,}\altaffilmark{3}}
          \author{Juan M. Uson\altaffilmark{4}}
          \author{Fran\c{c}oise Combes\altaffilmark{5}}

\altaffiltext{1}{Harvard-Smithsonian Center for Astrophysics, 
60 Garden Street, MS-42, Cambridge, MA 02138 USA; Electronic mail: 
lmatthew@cfa.harvard.edu}            
\altaffiltext{2}{Purple Mountain Observatory, Chinese Academy of
Sciences, 2~West Beijing Road, Nanjing 210008, Peoples Republic of China}
\altaffiltext{3}{Department of Astronomy, University of Massachusetts, LGRT B-619E, 710
N. Pleasant Street, Amherst, MA 01003 USA}
\altaffiltext{4}{National Radio Astronomy Observatory, 520 Edgemont Road
            Charlottesville, VA 22903 USA}   
	\altaffiltext{5}{LERMA, Observatoire de Paris, 61 Av. de l'Observatoire,
         F-75014, Paris, France}
\singlespace
\tighten

\begin{abstract}
Using the IRAM 30-m telescope, we have obtained 
$^{12}$CO $J$=1-0 and 2-1 spectral line observations
toward the nuclear regions of 15 edge-on, low surface brightness 
(LSB) spiral galaxies. Our
sample comprises extreme late-type LSB spirals with disk-dominated
morphologies and
rotational velocities $V_{\rm rot}\lsim$120~\kms. We report
detections  of
four galaxies   in at least one transition ($\gsim$5$\sigma$); for 
the remainder  of the sample we
provide upper limits on the nuclear CO content. 
Adopting a standard Galactic $I_{\rm CO}$-to-H$_{2}$
conversion factor implies molecular gas masses of
(3.3-9.8)$\times10^{6}~M_{\odot}$ in the nuclear regions (inner 1.1-1.8~kpc) 
of the detected galaxies. Combining our new data with samples
of late-type spirals from the literature, we find that the CO-detected
LSB spirals
adhere to the same $M_{\rm H_{2}}$-FIR correlation as more luminous
and higher surface brightness galaxies. The amount of
CO in the central regions of late-type spirals appears to depend more
strongly on mass than on central optical surface brightness, and
CO detectability declines significantly for moderate-to-low surface
brightness spirals with
$V_{\rm rot}\lsim$90~\kms; no LSB
spirals have so far been detected in CO below this threshold.
Metallicity effects alone are unlikely to
account for this trend, and we speculate that we are seeing
the effects of a decrease in the mean fraction of a galaxy disk 
able to support
giant molecular cloud formation with decreasing galaxy mass. 
\end{abstract}



\section{Introduction}
Late-type, low surface brightness (LSB) spirals are among the most
common disk galaxies in the local universe. These systems are
characterized by dim optical appearances (having central disk
surface brightnesses $\mu(0)_{B}\gsim$23 mag arcsec$^{-2}$), that reflect
both a low stellar surface density and a low global star formation
rate. While LSB spirals are found spanning a range of Hubble types, the
majority are  pure disk galaxies, 
with masses typically an order of magnitude smaller than an
$L_{*}$ spiral (McGaugh, Schombert, \& Bothun 1995; Sprayberry et al. 1997).
Numerous studies have shown that the bulk of these systems are
metal-poor and
rich in neutral hydrogen gas, implying they  
have remained inefficient
star formers over most of their lifetimes 
(Bothun, Impey, \& McGaugh 1997 and references therein).

Kennicutt's (1989) parameterization of the star formation efficiency
of disk galaxies predicts that
LSB spirals should be forming few new stars, since most or
all of their disks fall below the surface density threshold required
for efficient star formation (van der Hulst et
al. 1993; de Blok, McGaugh, \& van der Hulst 
1996; Uson \& Matthews 2003; see also Martin
\& Kennicutt 2001). 
Nonetheless, most late-type LSB
spirals do show signatures of modest ongoing star formation,
including blue optical colors (e.g., McGaugh \& Bothun 1994; Matthews \&
Gallagher 1997) and the presence of 
H$\alpha$ emission (e.g., Schombert et
al. 1990; McGaugh et al. 1995). In  a few
nearby LSB spirals, candidates for isolated supergiant
stars have also been identified 
(Gallagher \& Matthews 2002). Findings such as these have 
hinted that star formation in LSB
spirals may be governed by local conditions, or otherwise 
proceed in a rather different manner from the classic picture
of self-regulated star formation in giant molecular clouds (GMCs)
(see Schombert et al. 1990; Bothun et al. 1997; Gallagher \& Matthews
2002).

Overall,
we still understand relatively little about the physics of star formation in the
low-density, low-metallicity 
environments that pervade LSB disks. Similar conditions are also
expected to prevail in
a variety of other astrophysically interesting settings as well, including 
the outskirts of giant spirals (Ferguson 2002), 
moderate-redshift damped Lyman~$\alpha$ absorption systems (Chen et
al. 2004),
and in high-redshift protogalaxies.  
It is expected that in these types of environments, 
low ISM densities should increase dynamical timescales,  making it more
difficult to form and maintain molecular clouds; meanwhile, low metallicities
should decrease the cooling efficiency of the gas and result in fewer
grain surfaces on which H$_{2}$ molecules can form.

To date, most studies on the effects of low metallicity and
low ISM density on star formation have focussed on dwarf galaxies (see
Hunter 1997). Nonetheless,  
while LSB spirals share
some physical properties with typical dwarf irregulars
(e.g.,  subsolar metallicities; high fractional \HI\ masses), 
they tend to have thinner, more organized disk structures,  higher
rotational velocities, higher
dynamical masses, and rotation curves that are not purely solid
body. Many also show some degree of spiral structure. 
These factors may all influence the ISM
structure and resulting star formation efficiency through effects on
the amount of dynamical 
shear, gas scale height, effective intensity of the 
ultraviolet (UV) radiation field, the strength of the galactic
potential well, and the amount of pressure in the ISM (see Elmegreen
\& Parravano 1994; Hunter, Elmegreen, \& Baker 1998; Blitz \&
Rosolowsky 2004; Mac~Low \& Klessen 2004 and references therein). 
However, the interplay between all of
these variables in controlling molecule formation, 
ISM structure, and subsequent star formation in different galaxian
environments is still poorly
constrained, and it
has been debated how readily molecules and molecular clouds 
can form and persist in LSB spirals, 
and what would be the structure and distribution 
of their molecular gas component (Schombert et al. 1990;
Mihos, Spaans, \& McGaugh 1999; Gerritsen \& de Blok 1999; Mac~Low \& Klessen 2004).

One obvious means of obtaining insight into the above questions is through
direct searches for 
molecular gas in LSB galaxies.  Unfortunately, such
observations have proven to be challenging, and
initial efforts to detect molecular gas
in late-type LSB spirals through
CO spectral line observations failed to yield any detections (Schombert et
al. 1990; de Blok \& van der Hulst 1998). This led to
the suggestions that the
metallicities of typical LSB spirals may be too low for the formation
of CO molecules, or  that star formation in these systems may occur
directly from the atomic medium (Schombert et al. 1990; Bothun et
al. 1997). 
However, the latest-type spirals are in general  not strong CO
emitters (Young \& Knezek 1989; B\"oker, Lisenfeld, \& Schinnerer 2003; hereafter BLS03), 
and some of the initial CO observations of LSB spirals lacked the sensitivity needed
to establish whether for a given Hubble type, \HI\
mass, or dynamical mass, extreme late-type LSB spirals 
differ significantly in terms of molecular gas content
from their higher
surface brightness (HSB) counterparts.

More recently,
using observations $\sim$1.5-2$\times$ more sensitive 
than previous CO surveys of LSB spirals,
Matthews \& Gao (2001; hereafter MG01) detected $^{12}$CO(1-0) emission from
the central regions of 
three late-type, edge-on, LSB spiral galaxies. 
The observations of MG01 established
for the first time that at least some late-type LSB spirals contain modest
amounts of molecular gas, and moreover, that CO traces at least some
fraction of this gas. 

Five additional detections of LSB spirals have  been recently reported by
O'Neil et al. (2000,2003) and O'Neil \& Schinnerer (2004).
However, the detections of O'Neil and collaborators comprise giant, bulge-dominated 
systems with large rotational velocities ($W_{\rm 20,HI}/{\rm
sin}i>450$~\kms). 
These galaxies are  likely to have had a
rather different evolutionary history from the more common low-mass,
disk-dominated LSB galaxies (Hoffman, Silk, \& Wyse 1992; Noguchi
2001). In addition, the giant LSB spirals tend to have dense, metal-rich bulges 
(Sprayberry et al. 1995), which may significantly influence the
amount of molecular gas present in their inner regions. 
Because of these differences, we  limit our
discussion in the present paper to the molecular gas properties 
of disk-dominated, extreme late-type LSB spirals.

While the CO detections reported by MG01 are tantalizing,
the limited number of late-type LSB spirals so far surveyed 
in CO with good sensitivity has precluded establishing any firm
correlations between the presence of CO
and other properties of LSB disks, such as optical color, \HI\
content, far-infrared (FIR)
luminosity, rotation curve shape, global star formation rate,
or dynamical mass. To improve this situation, 
we have undertaken new observations of 15 extreme late-type
LSB spiral galaxies in the $^{12}$CO $J$=(1-0) and (2-1) lines.

\section{Sample Selection\protect\label{sample}}
For their pilot CO survey of late-type, LSB spirals,
MG01 specifically targeted systems oriented near edge-on to the
line-of sight. This was done
both to enhance the probability of detection (by increasing the projected column depth of
the molecular gas), and to facilitate complementary measurements
of the ISM at other wavelengths, including identification of dust clumps and
measurements of disk scale heights and vertical disk structure 
in the target galaxies.
We have continued to employ this strategy in the present survey.

For our targets we selected 15 nearby, edge-on 
LSB spirals observable with the IRAM 30-m telescope during our
allocated observing periods. Criteria for the selection of our targets included: 
(1) pure-disk morphology (with the exception of UGC~3137, which has a
small bulge); (2) inclination $i>80^{\circ}$; 
(3) radial velocity $V_{h}\lsim$2000~\kms; (4) previous detection in the \HI\
line with an integrated \HI\ line flux $\int S_{\rm HI}dV>1$~Jy~\kms; and 
(5) classification as an LSB spiral. Because accurately assessing 
intrinsic disk surface brightness can be difficult for edge-on galaxies, the LSB
classification was based on a combination of criteria in addition to 
low deprojected optical central
surface brightness:
(a) $L_{B}/L_{\rm FIR}\gsim$20 based on {\it IRAS} FIR 
measurements; 
(b) weak or undetected 20-cm radio
continuum (log$P_{\rm 20cm}\le20$~WHz$^{-1}$; cf. Hummel 1981); (c) a
global star formation rate
$\lsim$0.05$M_{\odot}$~yr$^{-1}$ based on integrated \HA, FIR, and/or
20-cm radio continuum luminosity; (d) lack of
a continuous dust lane or evidence for strong central dust absorption based on
the visual inspection of optical images. 
Several of the galaxies in our sample also have rather thin disks with
large disk axial ratios ($a/b>10$), another characteristic of intrinsically 
low-density, LSB disks (Gerritsen \& de Blok 1999; Bizyaev \& Mitronova
2002; Bizyaev \& Kajsin 2004).

Six of our targets
were previously observed in the $^{12}$CO(1-0) line
by MG01 using the NRAO 12-m telescope (UGC~711, UGC~2082, UGC~4148,
IC~2233, UGC~6667, UGC~9242). Among these,
UGC~2082 was detected by MG01. We also reobserved NGC~1560, for which Sage (1993) had 
previously reported a $^{12}$CO(1-0)
detection. 

Some properties of our target galaxies are summarized in
Table~1. 

\noindent {\it Column 1:} Galaxy name.

\noindent {\it Column 2:} Hubble type from NED.\footnote{The NASA/IPAC Extragalactic 
Database (NED) is operated by 
the Jet Propulsion Laboratory, California Institute of Technology,
under contract with the National Aeronautics and Space
Administration.} 

\noindent {\it Column 3:} Heliocentric recessional velocity measured from \HI\
spectral line observations, taken from the references in Column~19.

\noindent {\it Column 4:} Distances 
derived using the baryonic
Tully-Fisher relation of McGaugh et al. (2000), assuming
$H_{0}$=70 \kms~Mpc$^{-1}$. The $K_{s}$ magnitudes in Column~9 were
used for the distance determinations when available, otherwise the $B$
magnitudes in Column~8 were
used. We note the distances to UGC~290, UGC~711, and UGC~9760 (all derived 
from $B$ magnitudes)
are roughly a factor of two smaller than distances derived
from the Virgo infall model of Mould et al. (2000); this suggests that the adopted
internal extinction corrections for these cases
may be overestimated  (see explanation to Column~8).

\noindent{\it Column 5:} Optical size from NED.

\noindent{\it Column 6:} Galaxy inclination (from
LEDA\footnote{http://leda.univ-lyon1.fr/} unless otherwise noted).

\noindent{\it Column 7:} Position angle, measured east from north, taken from LEDA.

\noindent{\it Column 8:} Apparent
$B$ magnitudes taken from the Third Reference
Catalogue of Bright Galaxies (RC3; de Vaucouleurs et al. 1991). No RC3
magnitude was available for UGC~290, so we have adopted 
the value from van Zee (2000). All
$B$ magnitudes have been corrected  for 
foreground extinction according to Schlegel, Finkbeiner, \& Davis 
(1998) and for internal extinction following 
Tully et al. (1998; their Eq.~11). We caution that 
these internal extinction corrections may be overestimates
for LSB galaxies (see Bergvall \& R\"onnback 1995; 
Matthews \& Wood 2001), but more accurate estimates are
currently unavailable for our sample.

\noindent{\it Column 9:} Apparent $K_{s}$ magnitudes from the 2MASS 
survey.\footnote{http://irsa.ipac.caltech.edu/Missions/2mass.html} These are the
`Kron magnitudes'  from the 2MASS Extended Source
Catalogue where available; otherwise they are the 2MASS `total'
magnitudes (see Jarrett et al. 2003).  

\noindent{\it Column 10:}
Deprojected central surface brightnesses in the $B$ or $R$ band, taken
from the references in Column~19.

\noindent{\it Column 11:} Absolute $B$ magnitude.

\noindent{\it Column 12:} \HI\ linewidth measured at 20\% peak maximum, taken from
the references in Column~19.

\noindent{\it Column 13:} \HI\ masses
derived using the 21-cm fluxes from the references in Column~19, via the relation 
$M_{\rm HI}=2.36\times10^{5}\int SdV D^{2}~M_{\odot}$, where $\int
SdV$ is the integrated \HI\ line flux in Jy \kms\ and $D$ is the
galaxy distance in Mpc.

\noindent{\it Column 14:} Ratio of \HI\ mass to $B$-band luminosity.

\noindent{\it Column 15:} Dynamical
mass computed from 
$M_{\rm dyn}=2.326\times10^{5}rV^{2}(r)$, where $r$ is the galaxy
radius in kpc and $V(r)$ is
in \kms. We have assumed $V(r)=(W_{\rm 20,HI}-20)/2~{\rm sin}i$ 
and $r=1.5a_{\rm kpc}/2$,
where $i$ is the galaxy inclination and 
$a_{\rm kpc}$ is the linear diameter in kpc, derived from the
optical angular diameter, $a$, in Column~5.

\noindent{\it Column 16:}
FIR luminosity derived from {\it IRAS} data, following the prescription 
of Helou et al. (1988). We adopted the 60$\mu$m and 100$\mu$m
fluxes from NED when available; otherwise we used
the SCANPI 
tool\footnote{http://scanpi.ipac.caltech.edu} to derive these values, 
taking as the 60$\mu$m and 100$\mu$m fluxes the ``template
amplitudes'' determined from the noise-weighted
mean scans at 60 and 100$\mu$m. A ``probable signal'' (as defined in
the SCANPI Users Guide) could be
identified at 100$\mu$m for all but four galaxies and at 60$\mu$m for
all but five galaxies.

\noindent{\it Column 17:} H$\alpha$ luminosity (uncorrected for extinction)
derived using the \HA\ fluxes 
from the references 
in Column~19. 

\noindent{\it Column 18:} Twenty-centimeter continuum flux 
derived using images from the NRAO VLA Sky Survey (NVSS; Condon et
al. 1998). With the exception of NGC~1560, 
none of the galaxies were detected at 20cm, so
upper limits were  derived by 
determining the RMS noise  per beam,
$\sigma_{\rm 20cm}$, within  an elliptical aperture whose shape, location, and area, $A$, were 
defined by the optical size and position
angle of the galaxy, as given in Table~1 and the central coordinates
given in Table~2. 
Three-sigma upper limits were then computed as $F_{\rm
20cm}\le 3\sigma_{\rm 20cm}\times \sqrt{A/A_{b}}$, where $A_{b}$ is
the beam area. The 20-cm flux for NGC~1560 was
determined by integrating directly within an aperture of area $A$.

\noindent{\it Column 19:} Notes to the table.

\section{Observations and Data Reduction\protect\label{observations}}
We carried out $^{12}$CO $J$=1-0 and 2-1 spectral line observations of our
15 target galaxies using the IRAM 30-m telescope at
Pico Veleta, Spain during the following dates: 2001 August 24,29-31; 
2001 September 1,3; 2001 November 2-5; and 2002 June 5-10. Rest frequencies of the
two CO transitions are 115.271~GHz and  
230.538~GHz, respectively. Both transitions were observed simultaneously
in two polarizations (H and V) using a set of four SIS receivers. 
For the 115~GHz observations, the A100 and B100 receivers were used in
conjunction with the 1~MHz filter banks. 
Each set of filter banks was configured
into two units of 512 channels, one for each polarization, yielding
for each a
512~MHz bandpass, and a channel separation of $\sim$2.6~\kms. For the
230~GHz observations, the A230 and B230 receivers were used together with 
the autocorrelator backend, split into two
independent sub-bands, each with a 512~MHz bandwidth and 1250~kHz
($\sim$1.6~\kms) channels. During the
June run, one of the 115~GHz receivers was unavailable, limiting us to
only a single polarization at that frequency.

For each target galaxy, data were obtained at a
single pointing toward its optical center.
The beam size of the 30-m telescope is 21$''$ at 115~GHz and \as{10}{5} at
230~GHz (Kramer \& Wild 1994), corresponding to $\sim$0.2-1.8~kpc
and $\sim$0.10-0.92~kpc, respectively, at the distances of our
targets. For all of the targets, the beam was therefore significantly
smaller than the optical diameter of the galaxy (see Table~1).

Data were acquired in series of six-minute scans, each comprised of
twelve 30-s subscans.
To obtain the flattest possible baselines, 
the wobbler switching (nutating subreflector) observing mode was employed
with a frequency of 0.5~Hz.  Beam throws large enough to avoid
galactic emission in the off-beam were used (160$''$-240$''$, depending on the
angular size and parallactic angle of the source),
alternating between east and west
offsets for each consecutive subscan.
To further aid in distinguishing weak lines from possible baseline
structure or other
artifacts,  roughly two-thirds of the data for each source were obtained
with the recessional velocity of the source shifted by $\pm$50~\kms\ from the
center of the bandpass. 

Conversion from backend signal to antenna temperature ($T^{*}_{A}$)
was achieved using the standard `chopper
wheel' method
(see Kutner \& Ulich 1981; Kramer 1997). A calibration measurement 
of this type was performed after acquiring each new target and then 
after every other scan. The stability of 
the temperature scale of the 115~GHz receivers 
was also checked during the 2002 June run 
using observations of the line calibrator IRC+10216 
(Liechti et al. 1991) at frequencies corresponding to the redshifted
$^{12}$CO(1-0) lines in our galaxies. The IRC+10216 observations showed
the temperature scale to be stable and 
reproducible to within $\sim$15\%.
IRC+10216 did not provide any suitably strong lines for
calibration checks near the frequencies of our redshifted
$^{12}$CO(2-1) lines.

Pointing and focus were checked frequently throughout our observations. RMS
pointing errors were found to be $\sim3''$, consistent with the accuracy
expected after a pointing model update by the IRAM staff (Greve,
Panis, \& Thum 1996).

Our total effective integration times ranged from 1.2-10.6 hours per galaxy per
transition (Table~2). System temperatures in individual scans
ranged from  172-368~K (3-mm) and from 175-660~K  (1-mm, $T^{*}_{A}$
scale).

The spectra we present here have been converted
from the $T^{*}_{A}$ scale 
(the effective antenna temperature above
the atmosphere, as derived directly from the standard IRAM 30-m calibration 
procedure) to the $T_{\rm mb}$ scale (main
beam brightness scale; see Mauersberger et al. 1989; Kramer 1997).
This conversion is
achieved through the relation $T_{\rm mb}=T^{*}_{A}\times(F_{\rm
eff}/B_{\rm eff})$ where
$F_{\rm eff}$ is the forward efficiency and $B_{\rm eff}$ is the main
beam efficiency. Values for
these latter quantities 
were provided by the IRAM staff at the time of the observations:
$F_{\rm eff}$=0.95 at 115~GHz and 0.91
at 230~GHz; $B_{\rm eff}$=0.74 at 115~GHz and 0.52
at 230~GHz. 

Our data reduction was performed using the CLASS software. Each
individual spectrum (six minutes total integration in one polarization) was inspected,
and any bad channels or interference features were
blanked. Such features were generally intermittent and/or
limited to single channels; their excision therefore
had no significant effect on 
the baseline determination or the shape of any detected
spectral lines. For nine of the galaxy targets, 
some individual spectra (2-10\% of the
total) showed
obvious non-linear baseline structures and were
discarded. Next, a first-order baseline was 
fitted (excluding the central 250~\kms\ of the bandpass) 
and subtracted from each usable spectrum. Finally, all of the individual
spectra for each source were
aligned in velocity and summed using weights based on the square root
of the system temperature (after correction for the atmospheric opacity).
For inspection and display purposes, the composite spectra
were smoothed using a Hanning
function to resolutions of 10.4~\kms\ and 6.5-13.0~\kms\
for the $^{12}$CO(1-0) and (2-1) lines, respectively.
However, line parameters, uncertainties, and upper limits were all computed
from the unsmoothed spectra.

\section{Results\protect\label{results}}
\subsection{CO Line Parameters}
The averaged, smoothed $^{12}$CO(1-0) and (2-1)
spectra for all of our program galaxies are presented in
Figure~1. Based on these spectra,
we have detected four of our targets at $\gsim 5\sigma$: UGC~231, UGC~2082, and UGC~6667 
in both transitions, and UGC~3137 in $^{12}$CO(1-0) only. 

For the detected galaxies, the CO line parameters (integrated fluxes,
line widths and velocity centroids) were estimated
using single Gaussian fits to the unsmoothed spectra. These fits are
overplotted on the panels of Figure~1.
In general, our spectra have insufficient
signal-to-noise to distinguish Gaussian line shapes from more
complex line profiles.  Uncertainties for the integrated
line intensities were computed following Braine et al. (1993):

\begin{equation}
\delta I_{\rm CO}=\sigma_{\rm rms}\sqrt{\Delta V_{\rm CO}\delta_{c}}
\left[f\left(1-\frac{\Delta V_{\rm CO}}{W}\right)\right]^{-0.5} 
\end{equation}

\noindent where $\sigma_{\rm rms}$ is the RMS noise in the final averaged
spectrum in units of K, $\Delta V_{\rm CO}$ is the CO line width in \kms, 
$\delta_{c}$ is the channel spacing in \kms,
$f=\Delta V_{\rm CO}/\delta_{c}$, and $W$ is the
entire velocity coverage of the spectrum. This formula is designed to
take into account uncertainties both due to RMS temperature
fluctuations as well as baseline irregularities. Quoted uncertainties
for the line centroids and widths are based on the formal errors to
the Gaussian fits. Because of the low signal-to-noise of the spectra
and possible deviations from inherently 
Gaussian shapes, these values are expected to
underestimate the true uncertainties in the line widths by a factor of
$\sim$2-3, and the quoted line widths should be regarded only as estimates. 

In cases where no signal was present,
we computed a 3$\sigma$ upper limit to the CO flux in the nuclear
region of the galaxy using the relation

\begin{equation}
I_{\rm CO}\le3\sigma_{\rm rms}\sqrt{W_{20,\rm HI}\delta_{c}}
\left[f\left(1-\frac{W_{20,\rm HI}}{W}\right)\right]^{-0.5}
\end{equation}

\noindent where $W_{20,\rm HI}$ is the \HI\ line width at
20\% peak maximum (see Table~1) and $f=W_{\rm 20,HI}/\delta_{c}$. The resulting
values are summarized in Table~3.

\subsection{Molecular Hydrogen Masses}
Given the low densities and metallicities of LSB galaxies,
the appropriate
conversion factor (``$X$'') between the directly measured integrated CO line flux,
$I_{\rm CO}$,  and a total
mass of molecular hydrogen within the beam is rather uncertain
(see e.g., Mihos et al. 1999). 
Several authors have suggested prescriptions
for estimating $X$ based on galaxy luminosity (e.g., Wilson
1995; Israel 1997; Boselli, Lequeux, \& Gavazzi 2002), but the
recommendations
depend on a number of assumptions that are poorly constrained for
LSB spirals (e.g., the strength of the UV
radiation field; the adherence of LSB galaxies to standard
mass-metallicity relations). 
For simplicity, we 
therefore  adopt a standard Galactic 
$I_{\rm CO}$-to-H$_{2}$ conversion factor:
$X=N({\rm H_{2}})/\int T_{\rm mb}({\rm CO})dV=3.0\times10^{20}$~cm$^{-2}$~(K
km~s$^{-1}$)$^{-1}$ (Young \& Scoville 1991), where $\int T_{\rm
mb}dV=I_{10}$ [the integrated line flux of the $^{12}$CO(1-0) line]. 
We then estimate the total molecular hydrogen mass within the 21$''$
FWHP 
telescope beam as

\begin{equation}
M_{\rm H_{2}}=4.78[\pi/(4~\ln~2)]I_{10}d^{2}_{b} ~~(M_{\odot})
\end{equation}

\noindent where $I_{10}$ is the integrated $^{12}$CO(1-0) line flux in units
of K \kms\ ($T_{\rm mb}$ scale) and $d^{2}_{b}$ is the projected telescope beam
diameter in parsecs at the distance of the galaxy.  This formula
assumes a Gaussian telescope beam and that all the molecular gas resides in 
clouds in virial equilibrium. In Section~\ref{discussion} we 
briefly consider how  the possibility of a
metallicity-dependent $X$ value may impact our
results. For undetected galaxies, we
derived an upper limit to $M_{\rm H_{2}}$ (for the central regions of the
galaxy only) by
substituting into equation~3 the $I_{\rm CO}$ upper limits  derived from
equation~2. 
The resulting values of $M_{\rm H_{2}}$ for the nuclear regions of 
our four detected galaxies range from
$M_{\rm H_{2}}=(3.3-9.8)\times10^{6}~M_{\odot}$ (Table~3).

Because the 30-m beam sampled only the nuclear regions (inner
0.2-1.8~kpc) of our target galaxies, 
we are likely to have underestimated the {\it total} CO fluxes and
H$_{2}$ masses in most cases.
One of the LSB spirals we detected with the 30-m telescope (UGC~2082) was also
detected by MG01.
The value of $M_{H_{2}}$ derived by MG01 from a
single-pointing 12-m observation
(2.9$\pm0.5\times10^7 M_{\odot}$, adopting the distance from Table~1) is
a factor of 6 higher than the value from our 30-m measurement
(4.8$\times10^{6} M_{\odot}$), implying there is significant CO
beyond the 30-m beam. 
Consistent with this, the $^{12}$CO(1-0) line
measured by MG01 was considerably broader than the one measured from
the 30-m spectrum ($\Delta V_{10}|_{_{\rm 12m}}=176.9\pm26.4$ versus $\Delta
V_{10}|_{_{\rm 30m}}=57.5\pm7.5$), indicating that the larger 12-m beam 
sampled CO gas extending to larger physical
distances from the center of the galaxy (where rotational velocities
are higher). 

As another means of estimating how severely we may have underestimated the
total CO fluxes of our targets, we consider what is known
about the radial CO distributions of other late-type and/or LSB spirals.
In their large CO survey of field galaxies,
Young et al. (1995) found that
the radial extent of the CO emission relative to the optical diameter
has a mean value of $D_{\rm CO}/D_{25}\sim0.5\pm0.1$ for latest 
Hubble-type spirals (Scd-Sm). 
However, at 115~GHz, the 21$''$ beam of the 30-m telescope
subtended only 0.03-0.16$D_{25}$ in our LSB targets. 
While it is still uncertain whether the CO emission 
typically would be as extended in
LSB disks as in brighter late-type spirals, a comparison with major
axis mapping
measurements of two LSB spirals from MG01 suggests that the CO in LSB spirals is
in some cases extended over at least
the central $\sim$1-3~kpc of the galaxies
($\sim0.1-0.2D_{25}$). We conclude that in most cases, our present 
measurements should be regarded only as lower limits to the total CO fluxes.

An additional uncertainty that must be considered in estimating the
molecular gas masses of LSB spirals is the assumption that the
molecular gas in these galaxies resides in virialized clouds---the
physical basis for the use of the standard $X$ factor
(see Young \&
Scoville 1991). The adherence of LSB spirals to a standard $M_{\rm
H_{2}}$-FIR correlation and the presence of optically thick dark
clumps in some of these galaxies supports this assumption
(see below), although it is possible that overall a larger
fraction of the ISM of LSB galaxies is in a diffuse form rather than
dense clouds (Mihos
et al. 1999; Matthews \& Wood 2001). Finally, we note that for our
edge-on LSB sample, we have also assumed that the bulk of the gas clumps giving
rise to CO emission are not self-shadowed.

\subsection{Comments on Individual Galaxies\label{individual}}
{\bf UGC~231:} The velocity offset  between
the $^{12}$CO(1-0) and (2-1) line centers (26~\kms) is 
larger than expected from the formal uncertainties. We have verified
that both observations were taken with the 
correct redshifted frequency at the band center. One possibility is
that the larger $^{12}$CO(1-0) beam samples gas over a larger fraction of the
galaxy rotation curve than the $^{12}$CO(2-1) beam. 
This would be consistent with the larger width
of the $^{12}$CO(1-0) line. However, baseline structure in the (1-0) 
spectrum may also be a factor.

{\bf UGC~711:}  {\it Hubble Space Telescope} ($HST$) images reveal a few
patches of absorption in the inner disk
regions of UGC~711, but no evidence for an extensive dusty
component (Gallagher \&
Matthews, in prep.), consistent with the lack of  significant detected CO.

{\bf UGC~2082:} previously detected in the $^{12}$CO(1-0) line by MG01 with
the NRAO 12-m telescope. Our
new measurement yields a smaller value for $M_{\rm H_{2}}$, implying 
the molecular gas extends beyond the 30-m
beam. The broader width of the $^{12}$CO(2-1) line compared with the $^{12}$CO(1-0) line
may be caused by spurious baseline structure.

{\bf NGC~1560:} The emission features we detect in our $^{12}$CO(1-0) and
(2-1) spectra toward NGC~1560 are quite narrow (FWHM$\sim$1~\kms)
and are offset by +64~\kms\ from the \HI\ recessional velocity of
NGC~1560 (see Figure~\ref{fig:spectra}). 
Based on the \HI\ velocity field from
Broeils (1992), material is
present in NGC~1560 at these positive velocities only in the southern portion
of the galaxy, several arcminutes from our CO pointing
center. Additional test
observations we obtained 
offset by up to several degrees from NGC~1560 still show emission
present near $V_{h}$=+27~\kms, implying it is associated with
the Galactic foreground rather than with NGC~1560. 
In the Galactic all-sky CO map of Dame, Hartmann, \& Thaddeus 
(2001), no CO emission is visible directly along
the line-of-sight toward NGC~1560 ($l=138^{\circ}$, $b=+16^{\circ}$), 
although emission near $V_{h}\approx$27~\kms\
is observed at slightly lower Galactic latitudes ($b\lsim14^{\circ}$)
in this direction.
Sage (1993) reported a $^{12}$CO(1-0) detection of
NGC~1560 with the NRAO 12-m telescope at a velocity $V_{LSR}=-9$~\kms\
($V_{h}\approx-4.5$~\kms). The narrow feature he detected is 
offset in velocity both compared with our new 30-m spectrum
and with the recessional velocity of NGC~1560, suggesting it is also
Galactic in origin. Sage's pointing center was slightly offset
from ours ($\alpha_{2000}$=04$^{h}$ 32$^{m}$ 49.9$^{s}$,
$\delta_{2000}$=+71$^{\circ}$ 52$^{'}$ 52.4$^{''}$), but still
should have encompassed the same region sampled by the 30-m
beam; therefore the origin of the velocity offset between our
respective spectra is unclear.

{\bf UGC~3137:}  detected in the $^{12}$CO(1-0) line only. Lack of a
$^{12}$CO(2-1) signal makes this case somewhat tentative, hence further
observations are desirable.
This is the only galaxy in our sample that has a
small bulge component. Our observed CO linewidth is consistent with
the spread in \HI\ velocities observed in the central 21$''$ of the
galaxy by Garc\'\i a-Ruiz et al. (2002).

{\bf IC~2233:} A $\sim$4.5$\sigma$ feature is seen in our $^{12}$CO(2-1) spectrum of
IC~2233, but
we regard it as suspicious, since its width 
($\sim$95~\kms) is broader than the \HI\ linewidth in the central
21$''$ of the galaxy based on the VLA data of Matthews \& Uson (in
prep.). In addition, many of the individual $^{12}$CO(2-1) spectra 
for this galaxy show some baseline structure. IC~2233 appears to have
undergone a recent mild burst of star formation, as evidenced by its
very blue optical color ($B-R$=0.55) and
several bright \HII\ complexes (Matthews \&
Uson, in prep.). Matthews \& Uson find the \HI\ in IC~2233 has a 
rather clumpy distribution,
with the brightest \HI\ peaks corresponding closely 
with \HII\ complexes. 
Possible explanations for the lack of CO emission 
despite evidence for young stars and dense \HI\ gas are
that the most recent episode of star formation has temporarily depleted
the molecular gas supply, that existing molecular gas is not
centrally-concentrated, or that IC~2233
is extremely metal-poor. Based on the 2MASS data, 
IC~2223 has $J-K_{s}$=0.74. The
$B-V$ color from the RC3 ($B-V$=0.44) implies a metallicity of
$\sim$0.3$Z_{\odot}$ based on the model grid of Galaz et al. (2002;
their Figure~15). Such a value is typical of late-type, LSB spirals
(McGaugh 1994). 

{\bf UGC~6667:} MG01 reported a marginal $^{12}$CO(1-0) detection of this
galaxy. In our 30-m spectra, both the $^{12}$CO(1-0) and (2-1) lines show a velocity
offset of $\sim-25$~\kms\ compared with the published \HI\ recessional
velocity (Table~1), suggesting that our pointing center was displaced slightly from the
galaxy's kinematic center. The velocity widths of our detected lines
are consistent with the observed spread in \HI\ velocities near the
galaxy center based on the data from Verheijen (2001).
UGC~6667 has a rather red near-infrared (NIR)  color for a late-type LSB spiral 
($J-K_{s}$=1.23; Table~1), suggesting
it may be moderately metal-rich  (see Galaz et al. 2002).

{\bf UGC~8286:} Our limited integration time on this galaxy did not
permit very sensitive limits on its CO content. The similarity of its morphology,
rotational velocity, FIR, \HI, and \HA\ properties (Rand 1996; Swaters
et al. 2002) to those of
UGC~7321 (detected in CO by MG01) suggest further CO 
observations are desirable.

{\bf UGC~9242:}  UGC~9242 has a 
compact blue nucleus with starburst
characteristics  and a number of rather bright and compact \HII\ complexes visible
throughout its disk (Hoopes et al. 1999). The lack of detectable CO
emission from the nuclear regions of UGC~9242 therefore is somewhat
surprising; it may be partly linked to gas depletion or UV
dissociation of CO molecules from the nuclear
starburst, or to a low metallicity. The
$J-K_{s}$ color of UGC~9242 is 0.69 (Table~1); for
$B-V$=0.62 (RC3), Figure~15 of Galaz et al. (2002) implies a
metallicity of only $\sim$0.2$Z_{\odot}$. 

\section{Correlations between Molecular Gas Properties and Other
Physical Parameters of Extreme Late-Type Spirals\protect\label{discussion}}
Although the collective sample of late-type, LSB spirals for which deep CO
observations have been made is still rather limited,  here we 
explore some correlations in  the existing data between the
presence of CO and various other properties of  LSB and other extreme
late-type spiral galaxies. 

A valuable comparison sample for our study is
provided by the recent CO survey of extreme late-type
spirals by BLS03. 
Using the IRAM 30-m telescope, BLS03 surveyed the central regions of
an unbiased set of 47 bulgeless
spirals,  with $V_{h}\lsim$2000~\kms\ and Hubble types Scd-Sm. Their data do not go
as deep as ours (typically achieving RMS noise levels $\sim$2-3 times
higher at equivalent spectral resolutions),
but their observations sampled similar physical
scales in the target galaxies as our 30-m survey, and targeted
objects spanning a comparable range of disk structures, masses, and
Hubble types. The BLS03 survey is one of the first to provide CO
observations for a significant sample of late-type spirals
extending to objects of the lowest masses. By combining their
results with those from our present study, and from
MG01, we are able for the first time to examine the nuclear region
molecular gas properties of a set of low-mass, organized 
disk galaxies spanning a range
of optical surface brightness.

BLS03 tabulated face-on-corrected central surface brightnesses
in the $I$ band for all but seven  galaxies in their sample, and
Swaters \& Balcells (2002) published $R$-band photometry for three of the
remaining objects.  Assuming mean colors for 
late-type spirals (types 6$\le$T$<8$)  of
$B-I$=1.61$\pm$0.23 and $B-R$=1.06$\pm 0.26$ (de Jong 1996), 
and taking as a definition for LSB disks those with
a face-on
central surface brightnesses $\mu_{B,i}(0)\ge$23 mag arcsec$^{-2}$, 
this implies late-type,  LSB disks typically have
$\mu_{I,i}(0)\gsim$21.4~mag arcsec$^{-2}$ and
$\mu_{R,i}(0)\gsim$21.9~mag arcsec$^{-2}$. 
By these criteria, the BLS03 sample contains two LSB spirals [NGC~4204:
$\mu_{R,i}(0)$=21.9 mag arcsec$^{-2}$; and NGC~4395
$\mu_{R,i}(0)$=22.2 mag arcsec$^{-2}$]. NGC~4204 was detected in CO
by BLS03; NGC~4395 was not. The BLS03 sample spans a range in $I$-band
surface brightness of roughly 5 magnitudes and thus also includes a
number of galaxies of 
intermediate surface brightness. 

An interesting finding of BLS03 is that $\mu_{I,i}(0)\approx$18.7
mag arcsec$^{-2}$ marks a sharp division in CO brightness for
their sample (see their Figure~7). Among the galaxies with $\mu_{I,0}>18.7$ mag
arcsec$^{-2}$  only one of seventeen was detected
(to limits of $I_{\rm CO}\lsim$0.75 K~\kms). However, 
all 25 galaxies known to be brighter than $\mu_{I,i}(0)\approx$18.7
mag arcsec$^{-2}$ were detected.\footnote{One of these galaxies
(NGC~3906) was detected at $\sim3\sigma$ only in the $^{12}$CO(1-0)
line. With log$M_{\rm H_{2}}$=6.8 and $W_{\rm 20,HI}/{\rm sin}i$=69~\kms, it appears as an
outlyer on many of our plots, suggesting this detection may be spurious.} BLS03
suggested that this link between CO detectability and
optical surface brightness implies that the central mass density of
late-type spirals must be intimately linked with the molecular
gas content, even in the absence of a stellar bulge. As we discuss below,
our new results suggest that among moderate-to-low surface brightness disk galaxies,
the presence of detectable molecular gas may also depend on
other key physical parameters in addition to optical surface
brightness or central disk surface density.

In order to identify trends as a function of disk surface
brightness in the discussion that follows, we shall hereafter refer 
to those galaxies in the BLS03 sample with
$18.7\le\mu_{I,0}<21.4$~mag arcsec$^{-2}$  as
``intermediate surface brightness (ISB)'', and those 
with $\mu_{I,0}<18.7$ mag arcsec$^{-2}$ as 
``HSB'' extreme late-type spirals. We also collectively refer to the
LSB spirals in our new 30-m sample, together with those from MG01,
and the two LSB galaxies from BLS03, as
the ``combined LSB sample''. 
In the next section,
we now explore the CO properties of these extreme late-type spirals as a
function of various physical properties. 

\subsection{Molecular and Atomic Gas Masses of Extreme Late-Type
Spirals\protect\label{atomic}}
In Figure~\ref{fig:atomic} we plot nuclear molecular hydrogen mass as
a function of the total atomic gas mass for our combined LSB sample
and the extreme late-type spiral sample of BLS03.\footnote{We have rescaled
the  H$_{2}$ mass estimates
of BLS03 to account
for the different $X$ factor we have adopted here.}  Galaxies with
$M_{\rm H_{2}}$ upper limits are also shown\footnote{We do 
not include the upper limits derived by
MG01, since we have reobserved all of their undetected galaxies.}. Not
unexpectedly, 
there is some correlation between these two quantities in
the sense that the galaxies with the most atomic hydrogen also
tend to be the richest in molecular hydrogen in their centers. However, the
correlation is rather weak; galaxies with $M_{\rm
HI}\sim10^{8.5}-10^{10}~M_{\odot}$ are found with nuclear H$_{2}$
masses spanning a range of nearly three orders of
magnitude. Nonetheless, 
we see that the LSB and ISB objects tend to
have the smallest ratios of $M_{\rm H_{2}}/M_{\rm HI}$ for a given
\HI\ mass, consistent with their status as ``underevolved'' galaxies
(see also MG01).

It has often been argued that LSB spirals are inefficient star formers owing
to their low gas surface densities (e.g., van der Hulst et al. 1993;
de Blok et al. 1996). Therefore it is also of interest
to examine the relationship between the central H$_{2}$ surface
density, $\Sigma_{\rm H_{2}}$
(defined here as the nuclear molecular gas mass divided by the projected beam
area), versus the total, disk-averaged 
\HI\ surface density $\Sigma_{\rm HI}$ (the total \HI\ mass
divided by the area of the galaxy enclosed within the optical diameter
$D_{25}$). These quantities are plotted in
Figure~\ref{fig:atomicdens}.

We see that in general the LSB spirals in the combined sample fall at
the low end of the range of
$\Sigma_{\rm HI}$ values spanned by the entire extreme late-type sample, 
consistent with the low \HI\ surface densities
typically found in LSB spirals imaged in \HI\ with interferometers (e.g.,
van der Hulst et al. 1993; de Blok et al. 1996; Uson \& Matthews
2003). In contrast, the ISB and HSB subsamples show somewhat larger 
values of $\Sigma_{\rm HI}$. Moreover, 
the ISB and HSB galaxies show nearly identical
ranges of $\Sigma_{\rm HI}$ despite the fact that the HSB
galaxies on average have significantly higher nuclear
molecular gas surface densities 
than their ISB cousins. It is interesting then that 
several of the LSB spirals have been detected in CO, while many of the ISB
objects with higher disk-averaged \HI\ surface densities have not. 
These trends suggest that for  extreme
late-type spirals, $\Sigma_{\rm H_{2}}$
and $\Sigma_{\rm HI}$ are largely uncorrelated
over a range of roughly a factor of 10 in $\Sigma_{\rm HI}$. In
addition, higher optical central surface brightness seems to correlate
with larger central molecular gas density only for the highest surface
brightness galaxies. It would appear that some factor in addition to
the central baryonic density of galaxies thus plays a role in
governing the amount of molecular gas at their centers.

\subsection{The $M_{\rm H_{2}}$-FIR Correlation for LSB and Other Extreme
Late-Type Spirals\protect\label{FIR}}
For HSB spirals,
a correlation exists between the H$_{2}$ mass (or CO luminosity)
and the FIR luminosity (e.g., Young \& Scoville 1991
and references therein). 
This correlation presumably arises from the heating of
dust grains embedded in GMCs by young, massive stars (e.g., Telesco
\& Harper 1980). This relation might be expected to
break down for LSB spirals, if, for example, most of their molecular hydrogen resides
outside of GMCs, if they have a
bottom-heavy initial mass function (biased against OB star formation;
e.g., Lee et al. 2004),
or if the $X$ factor is considerably different for these galaxies.
It is therefore of 
interest to examine the FIR properties of our LSB sample in relation
to their CO properties.

BLS03 explored the $M_{\rm H_{2}}$-FIR correlation for
their extreme late-type spiral  sample, as well as for the more
luminous, earlier-type
spirals from the 30-m 
survey of Braine et al. (1993). BLS03 found the extreme late-type spirals to
continue the approximately linear $M_{\rm H_{2}}$-FIR correlation set by the
Braine et al. sample\footnote{We have rescaled
the ratio found by BLS03
to account for the different $X$ factor we have adopted here.}, 
with $L_{FIR}/M_{\rm H_{2},nuc}\approx31~L_{\odot}/M_{\odot}$. Note this
derived ratio uses the
total FIR luminosity of the galaxy, but
includes the molecular gas only in the nuclear regions; 
in comparison, typical values for this ratio are
$\sim$2-5$L_{\odot}/M_{\odot}$
for individual IR-bright
GMCs in the Milky Way  (Mooney \& Solomon 1988), and for samples of 
isolated spiral galaxies
where a total rather than nuclear CO flux is measured (e.g., Young \& Scoville 1991).

Among the galaxies in our combined LSB sample, all seven of
the CO-detected objects
were also detected in the FIR 
by {\it IRAS} at 60 \& 100$\mu$m.  Among the twelve
undetected LSB galaxies (eleven from our 30-m sample, one from BLS03), 
four were undetected by {\it
IRAS}.  In Figure~\ref{fig:FIR} we 
plot the FIR luminosities versus the 
values of $M_{\rm H_{2}}$ for the combined
LSB sample, as well as for those ISB and HSB galaxies in the BLS03 sample
that were detected by {\it IRAS}.
Values of $L_{FIR}$
for the BLS03 sample were computed using the {\it IRAS}
fluxes from the NED database or derived using 
SCANPI (see Section~\ref{sample}). BLS03 already showed that their
extreme late-type sample follows a fairly tight $M_{\rm
H_{2}}$-FIR correlation. Figure~\ref{fig:FIR} indicates
that additionally, all of the LSB spirals so far detected in CO 
adhere closely to the same  $M_{\rm H_{2}}$-FIR correlation as brighter
spirals.  This somewhat
surprising result tentatively suggests that as with brighter galaxies,
the CO we are detecting from LSB galaxies 
is associated with dusty GMCs heated by young massive stars, and
that the ``universal'' $X$ factor that we have adopted does not
result in a severe underestimate of their molecular hydrogen contents 
(at least compared with the nuclear regions of 
other galaxies of similar FIR luminosity). However, LSB galaxies
are typically metal-poor, and dust and CO molecules are both
comprised of heavy elements; therefore if the CO-to-dust ratio in most galaxies is fairly
constant,
this may also conspire to keep
the LSB galaxies on the same $M_{\rm H_{2}}$-FIR relation even if our
adopted $X$ value underestimates 
their nuclear H$_{2}$ contents.

Among those galaxies plotted on Figure~\ref{fig:FIR} that were  
detected by {\it IRAS} but undetected in CO, some of
the upper limits on the H$_{2}$ masses would still permit these
galaxies to adhere to a single  ``universal'' $M_{\rm H_{2}}$-FIR
correlation to within the expected scatter, although several likely
deviators are also seen.
A number of the LSB galaxies  that
are undetected in CO are  slow rotators
($V_{\rm rot}<90$~\kms), and as discussed below, these  are more
likely to be quite metal-poor and more likely 
to have an ISM structure containing few  GMCs. In these cases it
is unclear whether the observed FIR flux is linked with recent star formation,
or may instead arise from dust heated by late-type stars via
the interstellar radiation field.

\subsection{Molecular Gas Content and \HI\ Linewidth\protect\label{w20}}
In Figure~\ref{fig:w20} we plot the inclination-corrected \HI\
linewidths, $W_{\rm 20,HI}/{\rm sin}i$, 
versus the central molecular gas masses for the combined LSB
sample and the sample of  BLS03.  Values of $W_{\rm 20,HI}$
and $i$ for the BLS03 sample were taken from LEDA. From Figure~\ref{fig:w20}
we see that larger \HI\ linewidths correspond with
larger nuclear H$_{2}$
masses for extreme late-type spirals, with a
flattening in this relation reached near $W_{\rm 20,HI}/{\rm
sin}i\sim$250~\kms. We also see that 
among the late-type LSB spirals surveyed
in CO to date, no system with $W_{\rm 20,HI}/{\rm sin}i\lsim$200~\kms\ has
so far been detected, and
only a small number of galaxies in the BLS03 sample were detected below this 
threshold (and all are HSB objects). 
$W_{\rm 20,HI}/\sin i=200$~\kms\ corresponds
to a disk rotational velocity $V_{\rm rot}\approx$90~\kms, assuming
$V_{\rm rot}\approx(W_{\rm 20,HI} - 20)/2~{\rm sin}i$~\kms. Here a value of
20~\kms\ is used to correct for the contribution of the turbulent (non-rotational)
motions of the \HI\ gas to the total linewidth. 

The explanation for the trend in Figure~\ref{fig:w20} is not
immediately clear, as
a number of key galaxy properties are tied to rotational velocity,
including dynamical mass, depth of the galactic potential, 
and luminosity (as reflected in the Tully-Fisher
relation).  Mass and luminosity are in turn linked with
metallicity through the well-known luminosity-metallicity and mass-metallicity
correlations (see below). 
More slowly rotating 
galaxies also tend to have less rotational shear, which may be a key factor
affecting molecular cloud growth (e.g., Wyse 1986; Hunter et
al. 1998). 
Finally, Dalcanton, Yoachim, \& Bernstein (2004)
have recently reported an empirical link between peak galaxy rotational velocity and
the presence of dust lanes. Since all of these factors may 
affect the CO luminosities and/or molecular gas contents of galaxies,
in the subsections that follow,
we investigate some of these
properties more closely for the galaxies plotted in Figure~\ref{fig:w20}.

\subsection{Molecular Gas Properties as a Function of Dynamical Mass
and Optical
Luminosity}
Part of the effect seen in Figure~\ref{fig:w20} may be
a manifestation of the well-known mass-metallicity and luminosity-metallicity
relations for galaxies: lower-mass  (lower luminosity) galaxies in general have
lower mean metallicities (e.g., Bothun et al. 1984; Zaritsky,
Kennicutt, \& Huchra 1994;
Kobulnicky \&
Zaritsky 1999; Garnett 2002; Kuzio de Naray, McGaugh, \& de Blok 2004). 
This trend
is quite robust, and extends over a factor of $\sim$100 in metallicity and
$\sim$11 magnitudes in blue luminosity (Garnett 2002).
As discussed above, lack of metals
may affect observed CO luminosity in multiple ways, including
causing an underabundance of CO relative to H$_{2}$ and leading to a
dearth of dust grains to provide formation sites for molecules and 
shielding  from
dissociation by UV photons. Consistent with this, CO emission is
generally not
detected from the most metal-poor dwarf galaxies  (e.g.,
Taylor, Kobulnicky, \& Skillman 
1998).  To examine the possible role of metallicity effects on
the CO properties of extreme late-type spirals, 
we plot in Figure~\ref{fig:metalstest} the  nuclear
H$_{2}$ mass and nuclear H$_{2}$ surface density, respectively, 
versus blue absolute magnitude for the extreme late-type samples from
the previous figures. 

From Figure~\ref{fig:metalstest}a, we see that
lower-luminosity galaxies appear to have less
molecular gas
in their central regions than higher-luminosity galaxies (see also BLS03). 
We also see a drop-off in the detectability of CO near
$M_{B}\approx-16.5$. While it is
expected that more luminous galaxies should have more molecular gas
overall, since Figure~\ref{fig:metalstest}a plots only the nuclear H$_{2}$
contents within a finite beam, not the total molecular gas masses, 
a strong correlation is not necessarily expected. Since the lower luminosity
galaxies are predicted to  have lower metallicities, it is therefore possible
that we have significantly underestimated their nuclear H$_{2}$ contents by adopting
a Galactic $X$ value, leading to 
the apparent trend in Figure~\ref{fig:metalstest}a.
However, the empirical relationship between
$X$ and $M_{B}$ derived by Boselli et al. (2002) predicts that a decrease in blue
luminosity from  $M_{B}=-20$ to $M_{B}=-15$ should translate into an
increase in $X$ by only a factor of $\sim$8. In contrast, the mean nuclear
H$_{2}$ mass drops by more than two orders of magnitude over this $M_{B}$
range, suggesting that there are indeed real changes in the
concentration of molecular gas in extreme late-type spirals as a
function of optical luminosity (or equivalently, mass). 
Most of the lower-luminosity galaxies in
the extreme late-type sample are LSB or ISB systems, and 
there is some evidence that LSB galaxies produce lower
effective yields for a given luminosity than HSB galaxies (Garnett
2002). This would make them more metal-poor than  predicted by standard
luminosity-metallicity relations (see also Bergvall \& R\"onnback
1995; Bothun et al. 1997; Galaz
et al. 2002). However, it seems unlikely this would lead to any significant
revisions in $X$. Moreover, the relatively low UV
ionizing fluxes and comparatively high interstellar pressures expected
in some LSB disks relative to metal-poor dwarf galaxies (Matthews \&
Wood 2001) would act
to partially offset increases in  $X$ caused by a lack of metals (Mihos et al. 1999).

When we examine the relationship between
blue magnitude and $\Sigma_{\rm H_{2}}$
(Figure~\ref{fig:metalstest}b), we again see a
trend. Over the interval $-19\gsim M_{B}\gsim -17$ there is an apparent
decrease in the typical central molecular gas densities by 
nearly an order of magnitude.  This holds for galaxies spanning a
range in surface brightness. An analogous trend is also visible if we
examine $\Sigma_{\rm H_{2}}$ as a function of 
dynamical mass, $M_{\rm dyn}$, rather than blue absolute
magnitude (Figure~\ref{fig:mdyn}). Since substantial variations in $X$ are not
predicted over such a narrow mass or luminosity interval, it seems likely that
this result must stem in part from an intrinsic difference in the central gas surface
densities of galaxies over this range rather than from metallicity
effects alone. For the LSB galaxies, this
might be partly attributed to the decreased
efficiency of pure disk galaxies with low central matter densities in
concentrating molecular gas at their centers (see also
BLS03). However, we see that there is overlap in the 
$\Sigma_{\rm H_{2}}$ values  in the range $-17\lsim M_{B}\lsim -18$
for galaxies spanning a wide range in optical 
surface brightness. In the next section,
we speculate further on possible physical drivers for the trends seen in
Figures~\ref{fig:metalstest} \& \ref{fig:mdyn}.

\subsection{Molecular Gas Properties of Extreme Late-Type Spirals and
the ISM Structure of Low-Mass Spirals}
Dalcanton et al. (2004) recently uncovered an interesting trend
in a large sample of edge-on spiral galaxies: they found a 
sharp division near $V_{\rm rot}\approx$120~\kms\ in the
sense that
all galaxies with $V_{\rm rot}>$120~\kms\ exhibit dust lanes, while
no evidence for dust lanes is seen
in more slowly rotating galaxies. Dalcanton et
al. (2004) suggested that this transition reflects abrupt 
changes in the turbulent velocities 
supporting the gas layer, that are in turn governed by the onset of gravitational
instabilities in disks with $V_{\rm rot}\gsim$120~\kms. 
Predicted consequences are that the ISM in higher-mass galaxies would be prone to
fragmentation and efficient star formation, while lower-mass
disks would be inefficient star formers, and 
have more diffuse gas and dust with a larger scale height. Here we speculate
that the apparent falloff in the lack of detectable
CO emission for moderate-to-low surface brightness galaxies near
$V_{\rm rot}\lsim$90~\kms\ (Figure~\ref{fig:w20})
may also be linked to similar physical
processes that lead to the disappearance of well-defined dust lanes,
even though the CO cutoff manifests itself at somewhat lower rotational velocities.

UGC~7321 is one of the edge-on, LSB spiral galaxies detected in CO 
by MG01. Its rotational velocity
($V_{\rm rot}\approx105$~\kms) places it somewhat below the
boundary for the formation of disk instabilities proposed by Dalcanton
et al. (2004). However,
based on the analysis of optical {\it HST} images,
Gallagher \& Matthews (2002) reported evidence for two distinct
``zones'' within the galaxy, 
each exhibiting a rather different ISM structure and 
dominant mode of star formation (see also Matthews \& Wood 2001).
In the inner regions of the galaxy ($r<$2~kpc),
candidates for dark clouds consistent with GMCs 
are clearly seen, and young star-forming
regions appear to be confined to a very thin layer of scale height
$\sim$150~pc. In contrast, in the outer disk the dusty
clouds become less opaque and much
more diffuse, and candidates for individual supergiant stars
are visible well out of the midplane. Overall, the scale height of the
Population~I tracers is seen to increase with radius, consistent with
the radially increasing scale height of the \HI\ layer inferred by
Matthews \& Wood (2003). If the diffuse outer disk clouds are
analogous to Galactic diffuse clouds, then they are likely to be
primarily atomic rather than molecular (e.g., Elmegreen 1993).

Gallagher \& Matthews (2002) suggested that in the inner regions of
of LSB  spirals like UGC~7321, star formation
may be occurring in a classic, self-regulated mode, within well-defined GMCs,
while in the outer disk, it may be proceeding via locally-driven
processes (at a much
lower rate) in a medium largely supported by turbulence, 
where the gas scale height is no longer governed by star formation
processes (e.g., Sellwood \& Balbus
1999).  The boundary between
two such regions may depend upon rotational shear (Mac~Low \&
Klessen 2004), changes in the average gas pressure (Elmegreen \&
Parravano 1994; Blitz \& Rosolowsky 2004), and/or changes in the degree of self-gravity of the disk
(Sellwood \& Balbus 1999; Elmegreen \& Hunter 2000). 

The above findings, together with the results of our present CO survey,
suggest that
for galaxies in the mass range corresponding to 
$90\lsim V_{\rm rot}\lsim 120$~\kms, we
may be seeing the effects of a gradual decrease in the fraction of a
typical disk that is
self-regulated and unstable to molecular cloud and 
star formation. This, coupled with
metallicity effects, could explain the trend toward a 
falloff in detectable CO  among
systems in this mass range, and would  be consistent with the observed 
increase in the mean stellar and gaseous
scale heights relative to the radial extent of the disk 
for galaxies over this mass range  (Matthews \& van Driel
2000; Brinks, Walter, \& Ott 2002; Matthews \& Wood 2003).

\section{Summary\protect\label{summary}}
We have reported new $^{12}$CO(1-0) and $^{12}$CO(2-1) observations of
15 edge-on, late-type, LSB spiral galaxies. All of our targets
were nearby systems ($V_{h}\lsim$2000~\kms) with low to moderate masses
($V_{\rm rot}\lsim$120~\kms) and (with one exception) 
pure disk morphologies. We detected 
four of our target LSB galaxies 
in at least one CO transition ($\gsim5\sigma$). For the
remainder of the sample we report upper limits on the
CO content within the central regions. Our survey 
complements other recent CO surveys that have concentrated
on  giant, bulge-dominated LSB systems
with $V_{\rm rot}\gsim150$~\kms\ (O'Neil et al. 2000,2003; O'Neil \& Schinnerer
2004) or extreme late-type spirals of intermediate to high 
optical surface brightness (BLS03). 

The results of our study underscore that a bulge is not a prerequisite for
the presence of molecular gas in the central regions of LSB spiral
disks, and that at least some LSB spirals can support a multi-phase ISM.
Combining our results with those from the recent CO survey of 47
low-mass, late-type spirals by BLS03 
and the CO survey of late-type, LSB spirals by MG01, 
we find that the amount of CO in the nuclear regions
of the latest spiral types is not strictly a function of the central optical
surface brightness or the disk-averaged
\HI\ disk surface density; rather  lower mass galaxies in general tend
to have lower
central concentrations of molecular gas. Surprisingly, LSB spirals 
detected in CO follow the
same $M_{\rm H_{2}}$-FIR correlation as 
more luminous, HSB galaxies,  suggesting that the CO we are
detecting is tied to GMCs and current star formation
rather than to a more diffuse molecular medium.

We find that all LSB spirals and most
moderate-to-high surface brightness spirals that have been
detected in CO to date have $W_{\rm 20,HI}/{\rm sin}i\gsim200$~\kms\ 
(corresponding to
$V_{\rm rot}\gsim$90~\kms). The  fall-off in CO luminosity below
$V_{\rm rot}\gsim$90~\kms does not appear to be an artifact of limited observational
sensitivity. While lower-mass galaxies are expected
to be more metal-poor (which may result in CO becoming a poor tracer
of the total molecular gas content), we find that metallicity effects
alone are unlikely to fully explain this trend.
We speculate the drop-off in the detectability of CO in lower-mass
systems stems in
part from a decrease in the fraction of the disk that can support
self-regulated star formation; these changes
may be driven by changes in  the amount of rotational shear, changes in
average gas pressure or density, and/or changes in the degree of self-gravity of
the disk.
CO observations of larger samples of extreme late-type spirals,
coupled with optical imaging, surface photometry, and metallicity determinations are
needed to further explore these results.

\begin{acknowledgements}
We are grateful
to members of the IRAM staff for their support of this program,
and LDM and JMU acknowledge the valuable assistance from the Pico
Veleta staff during our observing runs.
LDM was supported by a Clay Fellowship from the Smithsonian
Astrophysical Observatory and the International Travel Grants
program of the National Radio Astronomy Observatory. 
YG is supported by the NSFC Nos. 10333060 \& 10425313.

\end{acknowledgements}

\clearpage

%
\begin{deluxetable}{lccllrrllclc}
\tabletypesize{\scriptsize}
\tablewidth{0pc}
\tablenum{1}
\tablecaption{Properties of the Target Galaxies}
\tablehead{
\colhead{Name} & \colhead{Type} &  \colhead{$V_{h,{\rm HI}}$} &
\colhead{$D$}
   & \colhead{$a\times b$} & \colhead{$i$} & \colhead{PA}
& \colhead{$m_{B}$}
& \colhead{$m^{T}_{K_{s}}$} & \colhead{$\mu_{\lambda,i}(0)$} &
\colhead{$M_{B}$} 
& \colhead{$W_{20,{\rm HI}}$} \\
     &  \colhead{}    & \colhead{(\kms)}   & \colhead{(Mpc)} &
\colhead{(arcmin)}    
& \colhead{($^{\circ}$)} 
& \colhead{($^{\circ}$)}  & \colhead{(mag)}
& \colhead{(mag)} &\colhead{(mag $''^{-2}$)} & \colhead{(mag)} &
\colhead{(\kms)}\\
\colhead{(1)}    &  \colhead{(2)}   & \colhead{(3)}   & \colhead{(4)}
& \colhead{(5)}    & \colhead{(6)} 
& \colhead{(7)}  & \colhead{(8)}
& \colhead{(9)} & \colhead{(10)} & \colhead{(11)} & \colhead{(12)}  }

\startdata
UGC231    &  Sc & 844 & 12.3 & 6.2\tm 0.6 & 89 &56 & 12.42& 10.52 & 26.7($B$) & -18.0
& 216\\

UGC290    &  Sdm & 767 & 5.3 & 2.2\tm 0.2 & 83 &136 & 15.76 &
... & 24.9($B$) & -12.9 &
98 \\

UGC711    &  SBd & 1982 & 12.8 & 4.6\tm 0.3 & 82 &118 & 12.85 & ... & ... &
-17.7 & 214 \\

UGC1281   & Sdm & 157 & 2.9 & 5.8\tm 0.6 & 84 &38 & 12.05 & ... & 23.9($B$) & -15.3 &
132 \\

UGC2082   & Sc & 710 & 10.6 & 5.9\tm 0.8 & 90 &133 & 11.86 & 10.04 & ... & -18.3
& 216\\

UGC2157   &  Sdm & 488 & 8.5 & 2.5\tm 0.6 & 90 &39 & 14.16 & 12.32 &...  &
-15.5 & 115\\

NGC1560   &  Sd & -37 & 2.2 & 11.6\tm 1.9 & 90 &23 & 10.71 & 8.78 &23.2($B$) & -16.0
& 155\\

UGC3137   &  Sbc & 992 & 17.4 & 4.6\tm 0.6 & 90 &74 & 13.30 & 10.76 &24.2($R$) &
-17.9 & 243 \\

UGC4148   & Sm & 737 & 8.3 & 2.5\tm 0.3 & 90 &10 & 14.37 & ... &.... & -15.2 &
141 \\

IC2233    &  SBd & 555 & 9.8 & 5.2\tm 0.6 & 87 &172 & 11.82 & 10.74 &22.5($R$) & -18.1 &
195\\

UGC4704   & Sdm & 599 & 6.3 & 4.1\tm 0.4 & 90 &115 & 14.19 & ... &... & -14.8 &
137\\

UGC6667   &  Scd & 973 & 16.5 & 3.7\tm 0.4 & 82 &89 & 13.09 & 11.67 & 23.8($B$) & -18.0 &
188 \\

UGC8286   &  Scd & 406 & 6.9 & 7.3\tm 0.8 & 90 &28 & 11.76 & 9.79 &20.9($R$) &
-17.4 & 189\\

UGC9242   &  Sc & 1439 & 18.2 & 5.4\tm 0.3 & 90 &71 & 12.57 & 11.84 &... & -18.7
& 208 \\

UGC9760   &   Sd & 2023 & 8.8 & 3.3\tm 0.2 & 90 &57& 13.35 & ... &... & -16.4
& 166\\
\enddata

\tablecomments{See Section~2 for a detailed explanation of the columns.}

\end{deluxetable}


%
\begin{deluxetable}{lllllcrr}
\tabletypesize{\scriptsize}
\tablewidth{0pc}
\tablenum{1}
\tablecaption{Properties of the Target Galaxies (cont.)}
\tablehead{
\colhead{Name} & \colhead{$M_{\rm HI}$}
& \colhead{$M_{\rm HI}/L_{B}$}
& \colhead{$M_{\rm dyn}$} &
\colhead{$L_{\rm FIR}$} & \colhead{$L_{H\alpha}$} &
\colhead{$F_{20{\rm cm}}$} 
& \colhead{Notes} \\ \colhead{} &
\colhead{($10^{9}M_{\odot}$)} &
\colhead{($M_{\odot}/L_{\odot}$)} &
\colhead{($10^{9}M_{\odot}$)} & 
\colhead{($10^{7}L_{\odot}$)} & \colhead{($10^{39}$erg s$^{-1}$)} &
\colhead{(mJy)}  & \colhead{} \\ \colhead{} &
\colhead{(13)} &
\colhead{(14)} & 
\colhead{(15)} & \colhead{(16)} & \colhead{(17)}  & \colhead{(18)} &
\colhead{(19)}  }

\startdata
UGC231  
& 1.5 & 0.60 &37. & 10. & 6.4 & $<$2.3& 1,2\\

UGC290  & 0.044 & 2.1 &0.90 &  ... & 0.09 &$<$1.1 & 2,3,4\\

UGC711   &  0.72 & 0.40 &28. & $<$4.6 & ... &$<$1.4 &  1\\

UGC1281   & 0.078 & 0.39 & 2.7 & 0.30 & 0.38 &$<$3.6 & 2,5\\

UGC2082   & 1.3 & 0.42 &30. &  11. & ... &$<$5.5 & 6\\

UGC2157   &0.041 & 0.11 &2.4 & ... & ... &$<$0.91  & 7\\

NGC1560   &0.42 & 1.1 &6.0 & 1.2 & ... & 14.3  & 8\\

UGC3137   &2.6 & 1.2 &50. & 23. & ... &$<$3.0 & 5,9\\

UGC4148   &0.20 & 1.0 &3.8 & ...& ... & $<$0.93 & 10\\

IC2233   &1.1 & 0.39 &20. & 6.6 & 5.8 &$<$3.0 & 9,11\\

UGC4704  &0.21 & 1.6 &4.5 & 1.0 & ... &$<$1.4 & 6\\

UGC6667   & 0.71 & 0.30 &22. & 7.5& ... & $<$2.3 & 12\\

UGC8286  &0.65 & 0.46 &18. & 3.2 & ... &$<$3.9 & 5,9\\

UGC9242  &1.5 & 0.31 &44. & 15. & 13.&$<$1.9 & 5,13\\

UGC9760  &0.27 & 0.50&7.8 & \nodata & \nodata &$<$0.87 & 3\\
\enddata

\tablecomments{See Section~2 for a detailed explanation of the
columns. Notes to Column 19:}
\tablenotetext{1}{\HI\ parameters from Lu et al. 1993}
\tablenotetext{2}{$\mu_{B,i}(0)$ and H$\alpha$
flux from van Zee 2000}
\tablenotetext{3}{\HI\ parameters from Giovanelli, Avera, \& Karachentsev 1997}
\tablenotetext{4}{$i$ and $m_{b}$ from van Zee 2001}
\tablenotetext{5}{\HI\ parameters from Garc\'\i a-Ruiz, Sancisi, \& Kuijken 2002}
\tablenotetext{6}{\HI\ parameters from Tifft \& Cocke
1988}
\tablenotetext{7}{\HI\ parameters from Matthews \& van Driel 2000}
\tablenotetext{8}{\HI\ parameters from Broeils \& van Woerden 1994; $\mu_{B,i}(0)$
    from Broeils 1992}
\tablenotetext{9}{$\mu_{R,i}(0)$ from Swaters \& Balcells 2002}
\tablenotetext{10}{\HI\ parameters from Bottinelli, Gougenheim, \& Paturel 1982} 
\tablenotetext{11}{\HI\ parameters and \HA\ luminosity from
Matthews \& Uson, in prep.}
\tablenotetext{12}{$\mu_{B,i}(0)$ and \HI\ parameters from Verheijen
2001}
\tablenotetext{13}{\HA\ luminosity from Hoopes, Walterbos, \& Rand 1999}

\end{deluxetable}

%
\begin{deluxetable}{lccrrccccl}
\tabletypesize{\scriptsize}
\tablewidth{0pc}
\tablenum{2}
\tablecaption{Summary of Observations}
\tablehead{
\colhead{Name} & \colhead{$\alpha$(J2000.0)} &
\colhead{$\delta$(J2000.0)}
 & \colhead{$t$(1-0)} &  \colhead{$t$(2-1)} &
\colhead{$T_{\rm sys}$(1-0)} &  \colhead{$T_{\rm sys}$(2-1)} & 
\colhead{RMS(1-0)} &  \colhead{RMS(2-1)} & \colhead{beam throw} \\
\colhead{}     & \colhead{($\ \ ^{h}$$\ \ ^{m}$ $\ \ ^{s}$)} 
& \colhead{($\ \ ^{\circ}$ $\ \ ^{'}$ $\ \ ^{''}$)} & \colhead{(min)} &
\colhead{(min)} & \colhead{(K)}    & \colhead{(K)}  
&  \colhead{(mK)}    & \colhead{(mK)} & \colhead{(arcsec)} \\
  \colhead{(1)} & \colhead{(2)} & \colhead{(3)} &
\colhead{(4)} & \colhead{(5)}  & \colhead{(6)} &
\colhead{(7)}  
&  \colhead{(8)}   & \colhead{(9)} & \colhead{(10)}}

\startdata
UGC231    & 00 24 02.8 & +16 29 10.0 & 156 & 150 & 248 &414 & 4.6   &5.9 & 240 \\

UGC290    & 00 29 08.5 & +15 53 57.0 & 132 & 258 &302 &516 & 4.8 & 5.3 & 200 \\

UGC711    & 01 08 37.3 & +01 38 25.9 & 240 & 230 &311 &761 & 5.1  &8.7 & 240 \\

UGC1281   & 01 49 31.7 & +32 35 16.2 & 144 & 144 & 278&481 & 4.8  &6.2 & 240 \\

UGC2082   & 02 36 16.3 & +25 25 24.0 & 188 & 170 & 319&492 & 5.9  &6.5 & 240 \\

UGC2157   & 02 40 25.3 & +38 33 46.2 & 144 & 138 &308&541 & 6.2 &8.9 & 240 \\

NGC1560   & 04 32 47.7 & +71 52 45.5 & 228 & 138 &390 &764 & 5.5   &11.7 &  240 \\

UGC3137   & 04 46 15.5 & +76 25 06.4 & 348 & 438 &353 &676 & 4.2  &6.2 & 240  \\

UGC4148   & 08 00 23.8 & +42 11 39.2 & 264 & 252 &257 & 441 & 3.6  &5.4 & 240 \\

IC2233    & 08 13 59.0 & +45 44 38.0 & 288 & 288 &284 &523 & 3.6 &5.1 &  240 \\

UGC4704   & 08 59 00.3 & +39 12 35.7 & 252 & 354 &296 &623 & 4.3  &5.7 & 200,240  \\

UGC6667   & 11 42 25.4 & +51 35 50.6 & 540 & 636 & 284  &450 & 2.8   &3.1 & 160,240 \\

UGC8286   & 13 12 11.8 & +44 02 16.0 & 72  & 138 & 310  &551 & 7.8 &9.1 & 200 \\

UGC9242   & 14 25 20.6 & +39 32 19.4 & 221 & 204 & 317  &642 & 5.3 &7.5 & 240 \\

UGC9760   & 15 12 02.3 & +01 41 52.1 & 168 & 168 &275 &583 & 5.0  &8.5 &  240   \\

\enddata

\tablecomments{
Explanation of columns: (1) Galaxy name; (2) \& (3) right ascension
and declination of the
pointing center (adopted from Cotton, Condon, \& Arbizzani 1999); 
(4) \& (5) total effective integration time, in minutes, on the
$^{12}$CO(1-0) and (2-1) lines; (6) \& (7) mean system
temperature (in K, on the $T_{\rm mb}$ scale) of the averaged $^{12}$CO(1-0)
and (2-1)
spectra; (8) \& (9) RMS noise (in mK) of averaged (unsmoothed) $^{12}$CO(1-0)
and (2-1) spectra;
(10)
beam throw, in arcseconds, used for the off-source observations (see Text).}
\end{deluxetable}

%
\begin{deluxetable}{lcccccccc}
\tabletypesize{\footnotesize}
\tablewidth{0pc}
\tablenum{3}
\tablecaption{Nuclear CO Properties of the Sample} 
\tablehead{\colhead{Name} & \colhead{$I_{10}$} &  \colhead{$I_{21}$} &
\colhead{$V_{10}$} 
& \colhead{$V_{21}$} & 
\colhead{$\Delta V_{10}$} & \colhead{$\Delta V_{21}$} &
\colhead{$M_{\rm H_{2}}$} 
& \colhead{$M_{\rm H_{2}}/M_{\rm HI}$} \\
 \colhead{}    &  \colhead{(K \kms)} &  \colhead{(K \kms)} &
\colhead{(\kms)} 
& \colhead{(\kms)} & \colhead{(\kms)} & \colhead{(\kms)} &
\colhead{($10^{6}M_{\odot}$)} &  \colhead{} \\
  \colhead{(1)} &  \colhead{(2)} & \colhead{(3)} & \colhead{(4)} &
\colhead{(5)} & \colhead{(6)}  & \colhead{(7)} & \colhead{(8)} & \colhead{(9)} } 

\startdata
UGC231    & 0.59$\pm$0.09 & 0.53$\pm$0.09 & 834.1$\pm$6.3 &
860.1$\pm$4.5 & 77.5$\pm$11.3 & 56.6$\pm$10.9 & 5.0 & 0.003 \\

UGC290    & $<$0.24 & $<$0.22 & ... & ... & ... & ... & $<$0.38 &
$<$0.009 \\

UGC711    & $<$0.39 & $<$0.59 & ... & ... & ...& ...& $<$3.6 & $<$0.005 \\

UGC1281   & $<$0.28 & $<$0.30 & ... & ... & ... & ... & $<$0.13 &
$<$0.002 \\

UGC2082   & 0.77$\pm$0.10 & 1.0$\pm$0.1 & 698.9$\pm$4.0 &
721.7$\pm$5.9 & 57.5$\pm$7.5 & 94.7$\pm$11.8 & 4.8 & 0.004 \\

UGC2157   & $<$0.33 & $<$0.40 & ... & ... & ... & ... & $<$1.3 & $<$0.032 \\

NGC1560   & $<$0.35 & $<$0.63 & ... & ... & ... & ... & $<$0.10 &
$<$0.0002 \\

UGC3137   & 0.58$\pm$0.1 & $<$0.46 & 1006.6$\pm$6.0 & ... &
74.5$\pm$14.7 & ... & 9.8 & 0.004\\

UGC4148   & $<$0.22 & $<$0.27 & ... & ... & ... & ...& $<$0.85 &
$<$0.004 \\

IC2233    & $<$0.27 & $<$0.32 & ... & ... & ... & ... & $<$1.4 & $<$0.001 \\

UGC4704   & $<$0.27 & $<$0.28 & ... & ... & ... & ... & $<$0.57 &
$<$0.003 \\

UGC6667   & 0.22$\pm$0.05 & 0.18$\pm$0.04 & 946.8$\pm$6.9 &
948.4$\pm$3.9 & 56.6$\pm$15.4 & 39.0$\pm$7.9 & 3.3 & 0.005 \\

UGC8286   & $<$0.56 & $<$0.56 & ... & ... & ... & ... & $<$1.5 &
$<$0.002 \\

UGC9242   & $<$0.40 & $<$0.49 & ... & ... & ... & ... & $<$7.5 &
$<$0.005 \\

UGC9760   &  $<$0.33 & $<$ 0.48 & ... & ... & ... & ... & $<$1.4 & $<$0.005 \\

\enddata

\tablecomments{All line parameters for detected galaxies 
are based on single-Gaussian fits. Upper limits
are derived using equation~2. Uncertainties quoted for the velocity widths of the
detected lines are the formal errors from Gaussian fits; actual
uncertainties are expected to be $\sim$2-3 times larger owing to the 
limited signal-to-noise of the spectra and possible non-Gaussian
shapes of the lines.
Explanation of columns: (1) Galaxy name; (2) \& (3) 
integrated line flux of
the $^{12}$CO(1-0) and (2-1) lines; (4) \& (5) line centroids 
of the $^{12}$CO(1-0) and
(2-1) lines; (6) \& (7) full width at
half-maximum velocity of the $^{12}$CO(1-0) and (2-1) lines;
(8) molecular hydrogen
mass within the central 21$''$ of the galaxy, estimated using
Equation~3; (9) ratio of molecular
hydrogen mass within the central 21$''$ of the galaxy to
the total atomic hydrogen mass.}
\end{deluxetable}


\begin{figure}
\vspace{-18.0cm}
\plotfiddle{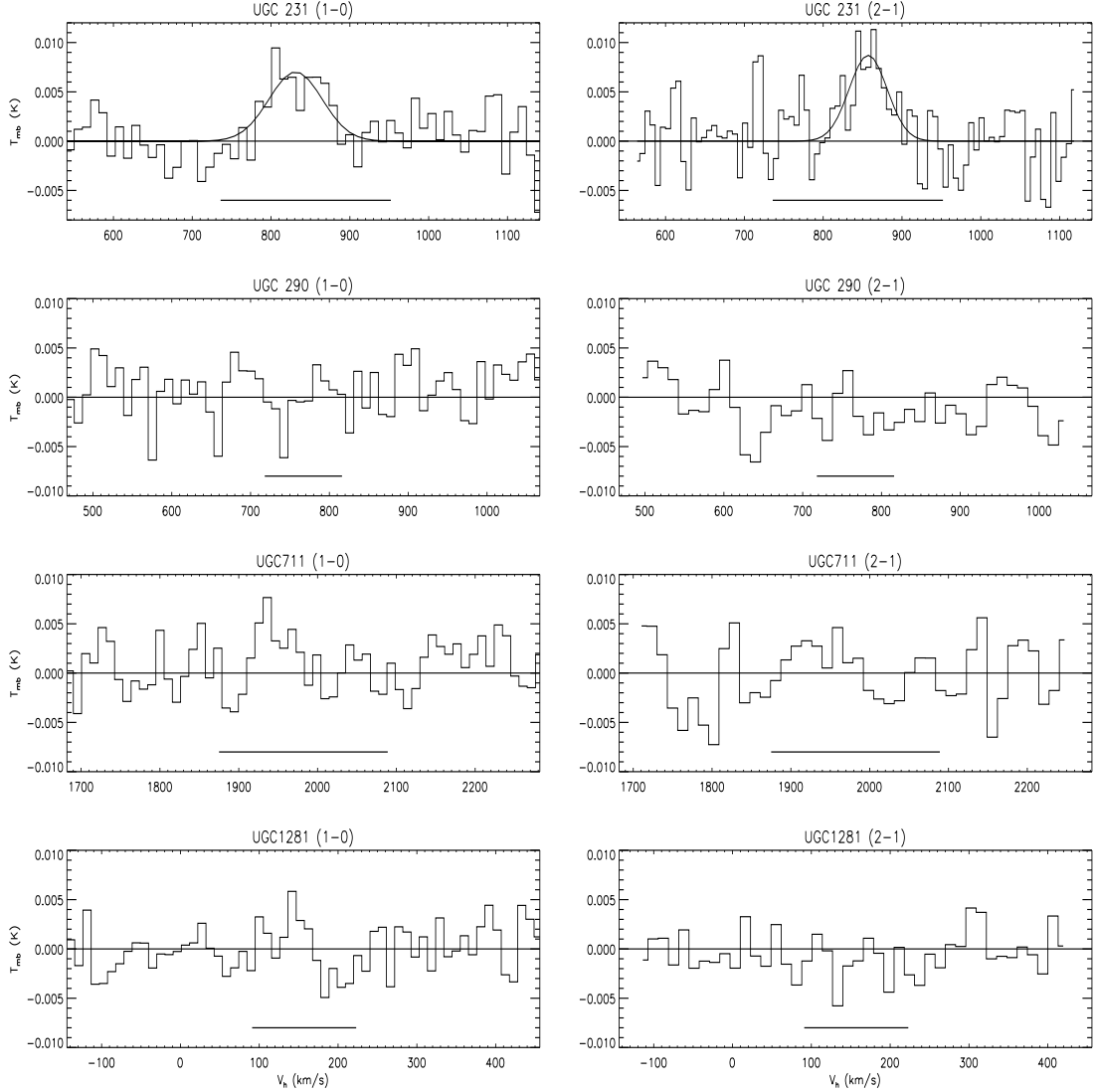}{7.0in}{90.0}{420}{420}{20}{350}

\caption{$^{12}$CO(1-0) and $^{12}$CO(2-1) spectra for the late-type, LSB
spirals observed in the present study. Axes are heliocentric radial
velocity, $V_{h}$, in \kms, and main beam brightness temperature, $T_{\rm mb}$,
in K. For display purposes, 
the spectra have been smoothed to resolutions of
10.4~\kms\ and 6.5-13.0 ~\kms, respectively (with the exception of the two
NGC~1560 spectra labeled ``full resolution'', that are shown at the observed
resolutions of $\sim$2.6~\kms\ and $\sim$1.6~\kms, respectively). 
The horizonal bars on each panel
denote the width of the \HI\ line at 20\% peak maximum,
$W_{\rm 20,HI}$, based on the references given in Table~1. For detected
galaxies, we have overplotted 
the Gaussian fit from which the CO line parameters were derived. All spectra
comprise a single ponting toward the nuclear regions of the 
galaxy.\label{fig:spectra}}
\end{figure}

\addtocounter{figure}{-1}
\begin{figure}
\vspace{-18.0cm}
\plotfiddle{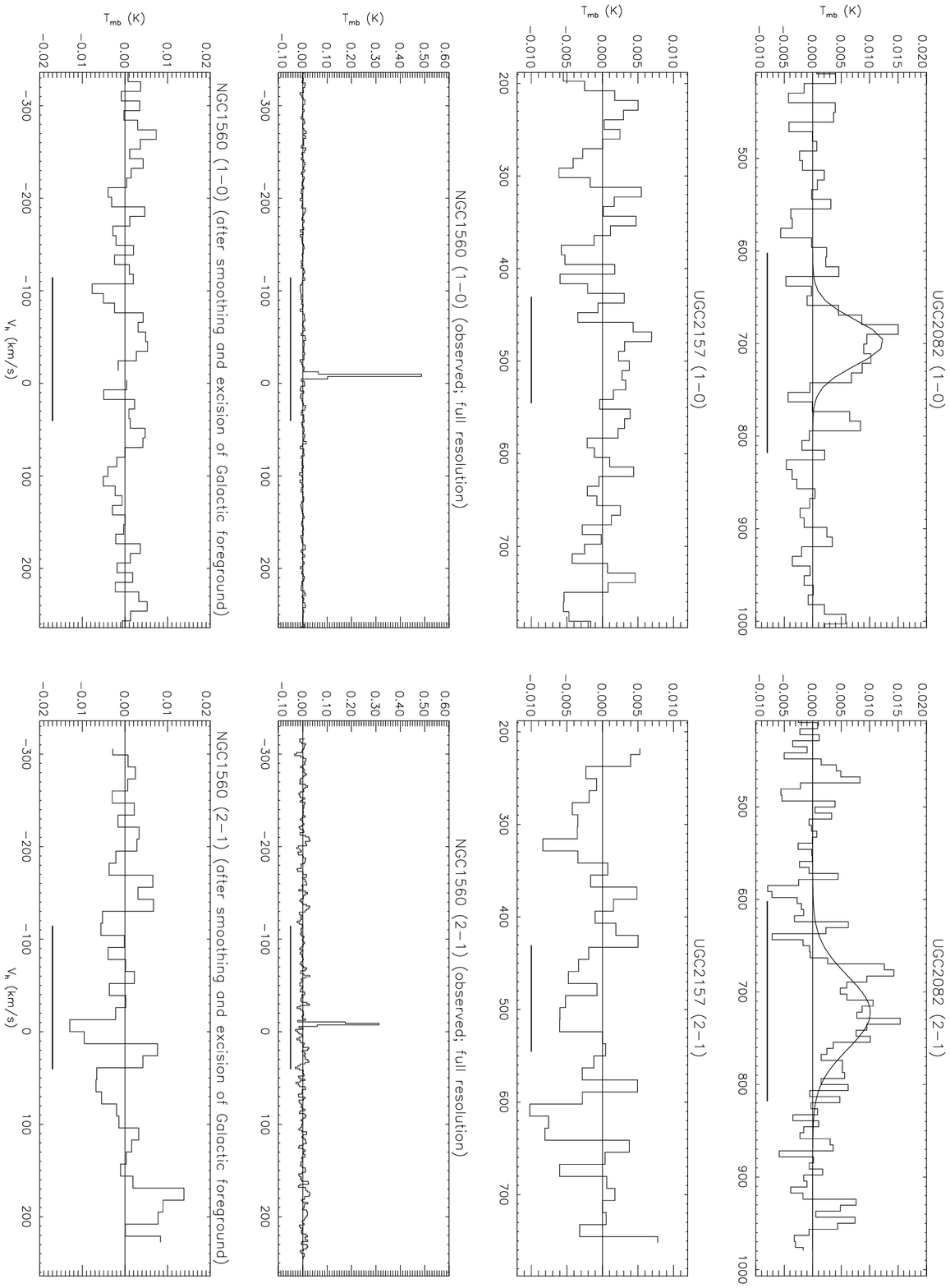}{7.0in}{90.0}{420}{420}{20}{350}
\caption{cont.}
\end{figure}

\addtocounter{figure}{-1}
\begin{figure}
\vspace{-18.0cm}
\plotfiddle{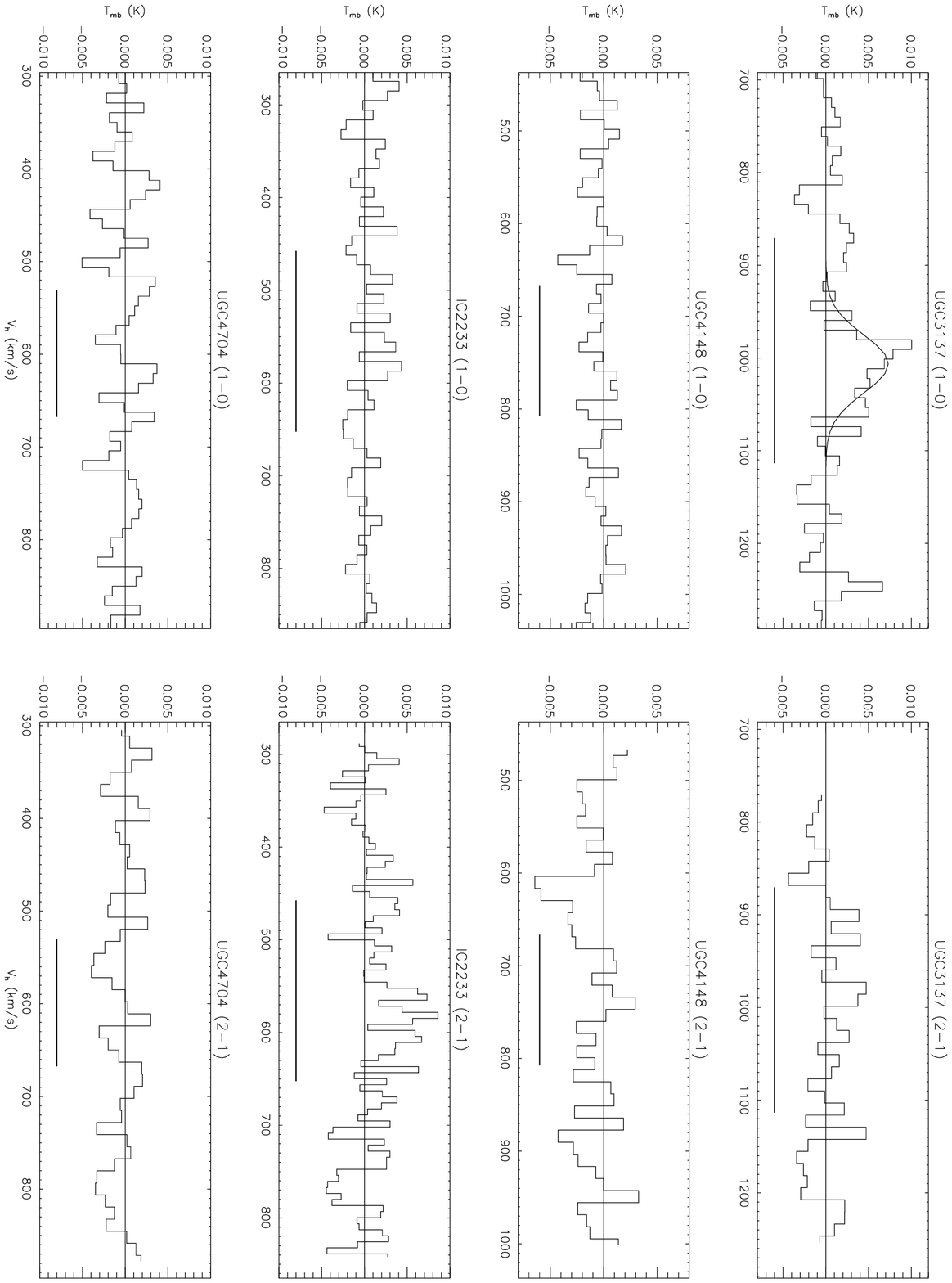}{7.0in}{90.0}{420}{420}{20}{350}
\caption{cont.}
\end{figure}

\addtocounter{figure}{-1}
\begin{figure}
\vspace{-18.0cm}
\plotfiddle{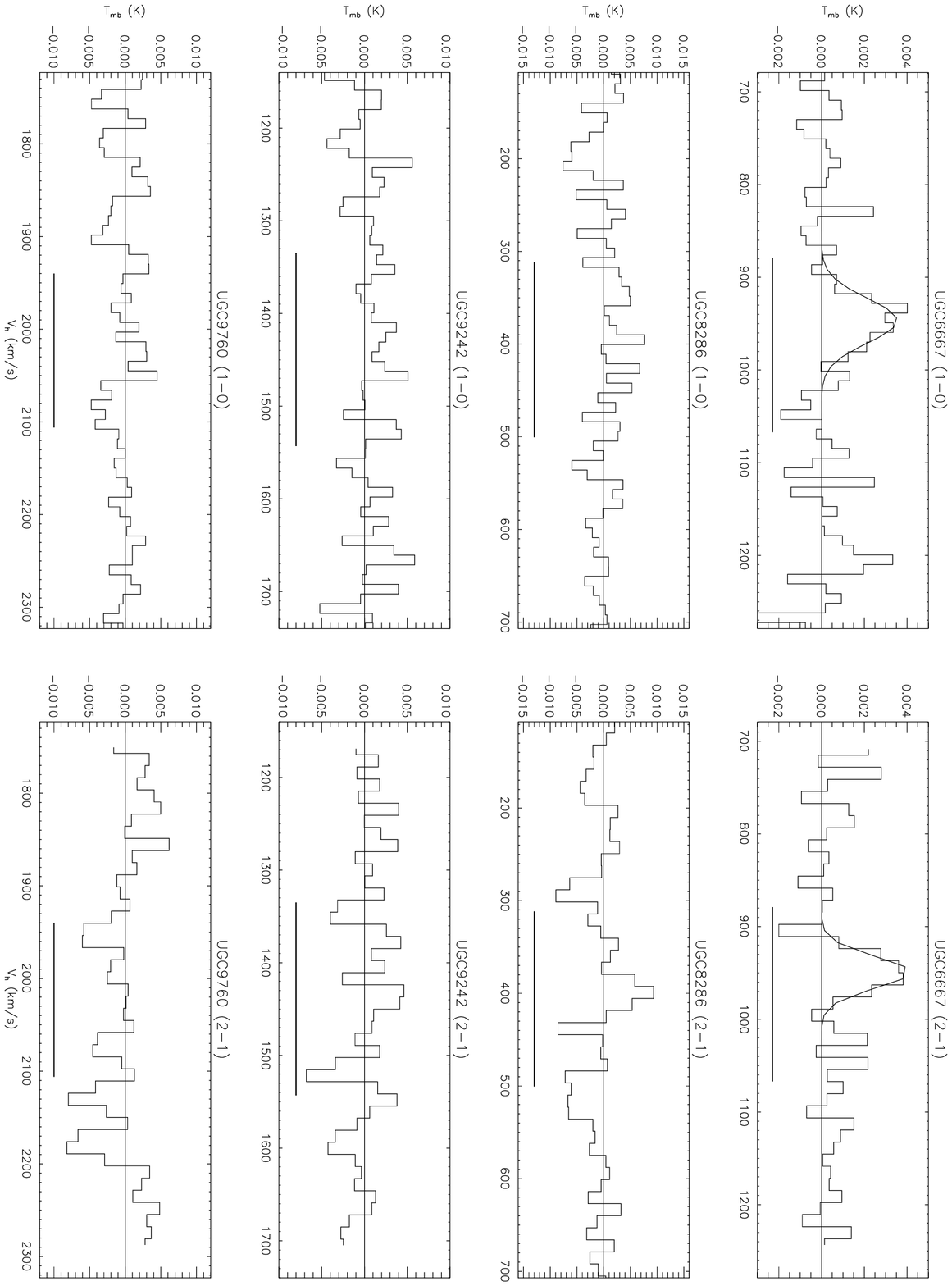}{7.0in}{90.0}{420}{420}{20}{350}
\caption{cont.}
\end{figure}

\begin{figure}
\vspace{-16.0cm}
\plotfiddle{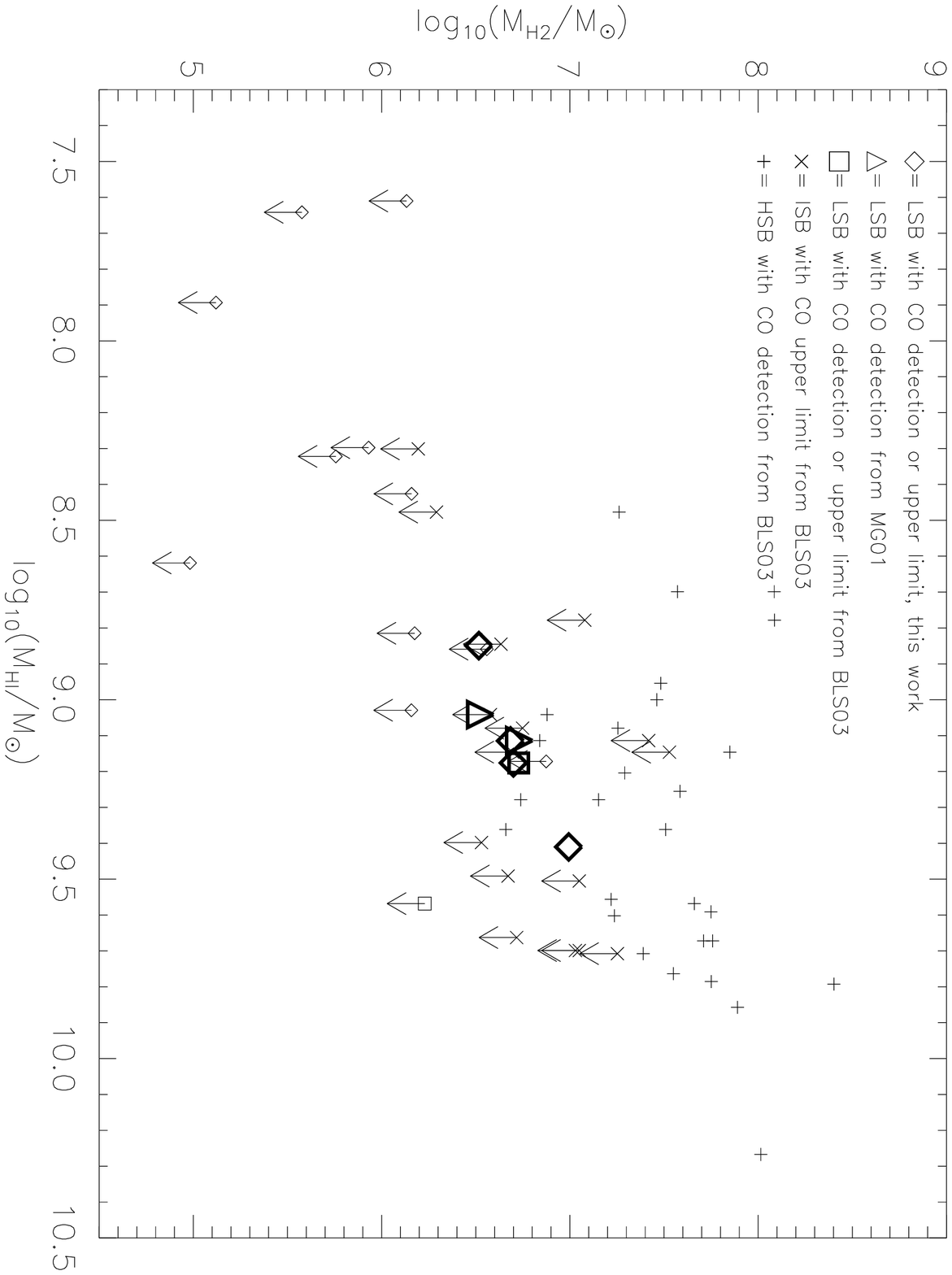}{7.0in}{90.0}{300}{400}{20}{350}
\caption{Logarithm of the total atomic hydrogen mass versus the
logarithm of the nuclear
molecular hydrogen mass (both in solar masses) for extreme late-type
spiral galaxies. LSB spirals observed as part of the present
study are shown as diamonds (15 galaxies); those from BLS03 as
squares (2 galaxies); and those from MG01 as triangles (2 galaxies); 
large bold symbols indicate
detected LSB galaxies ($\gsim5\sigma$) and smaller symbols denote cases
with upper limits. Additional galaxies from BLS03 
are shown as `+' symbols (for HSB objects) or `X' symbols (ISB
objects; see Text for definitions). All H$_{2}$ masses were derived
from data from the IRAM 30-m telescope ($21''$ FWHP beam at 115~GHz), with the
exception of the
points from MG01 (derived from the NRAO 12-m telescope; 55$''$ FWHP
beam). The $M_{\rm H_{2}}$ value for the MG01 point corresponding to UGC~7321 has
been scaled by the ratio of the 30m-to-12m beam areas to make the
subtended beam area
commensurate with the rest of the LSB sample.}
\label{fig:atomic}
\end{figure}

\begin{figure}
\vspace{-16.0cm}
\plotfiddle{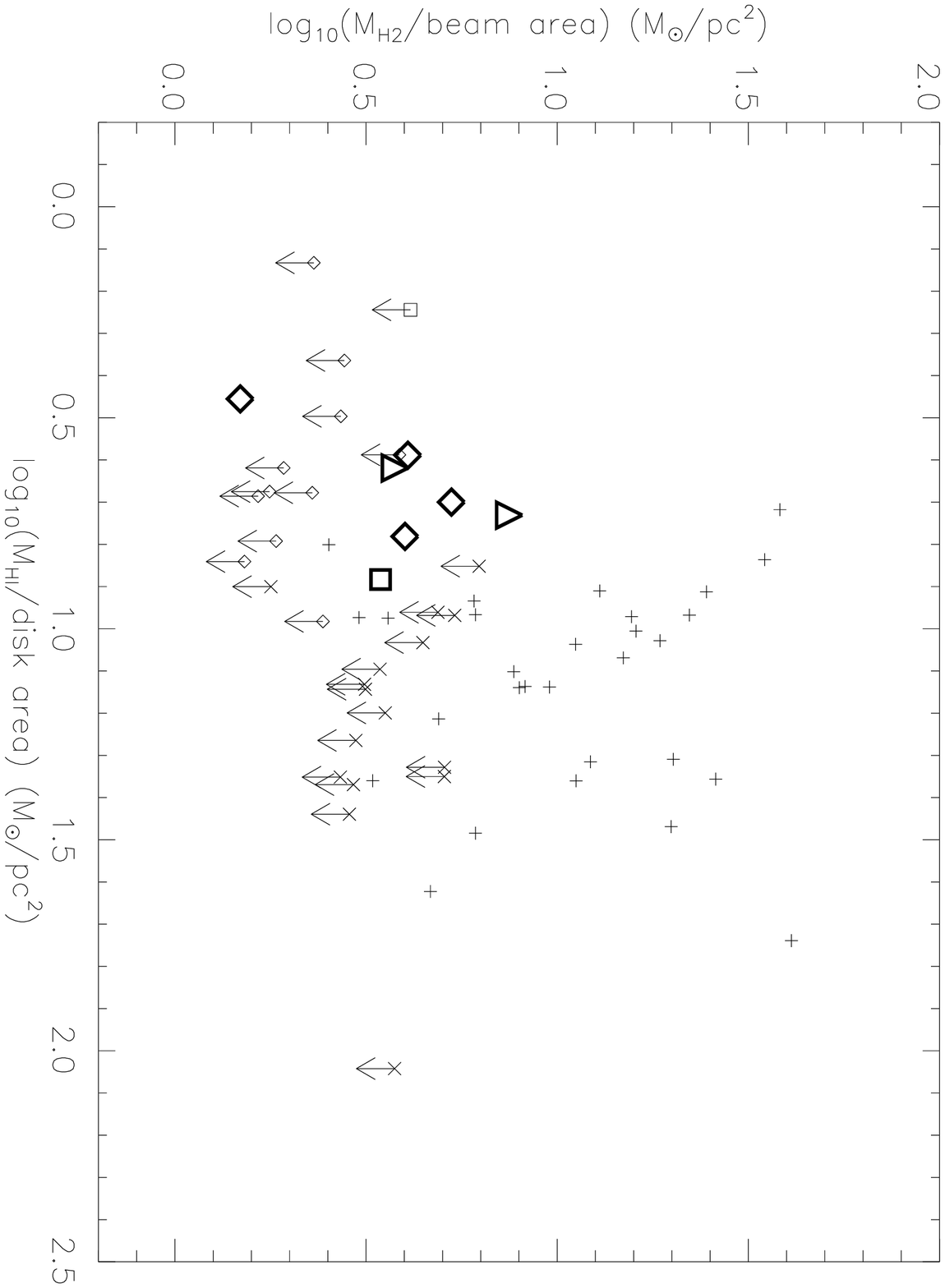}{7.0in}{90.0}{300}{400}{20}{350}
\caption{Logarithm of the mean \HI\ surface density
(averaged over the stellar disk) versus the logarithm of the
beam-averaged central H$_{2}$ density, both in solar masses per square
parsec for the combined LSB spiral sample and the extreme late-type
spiral sample of Figure~\ref{fig:atomic}.}
\label{fig:atomicdens}
\end{figure}


\begin{figure}
\vspace{-16.0cm}
\plotfiddle{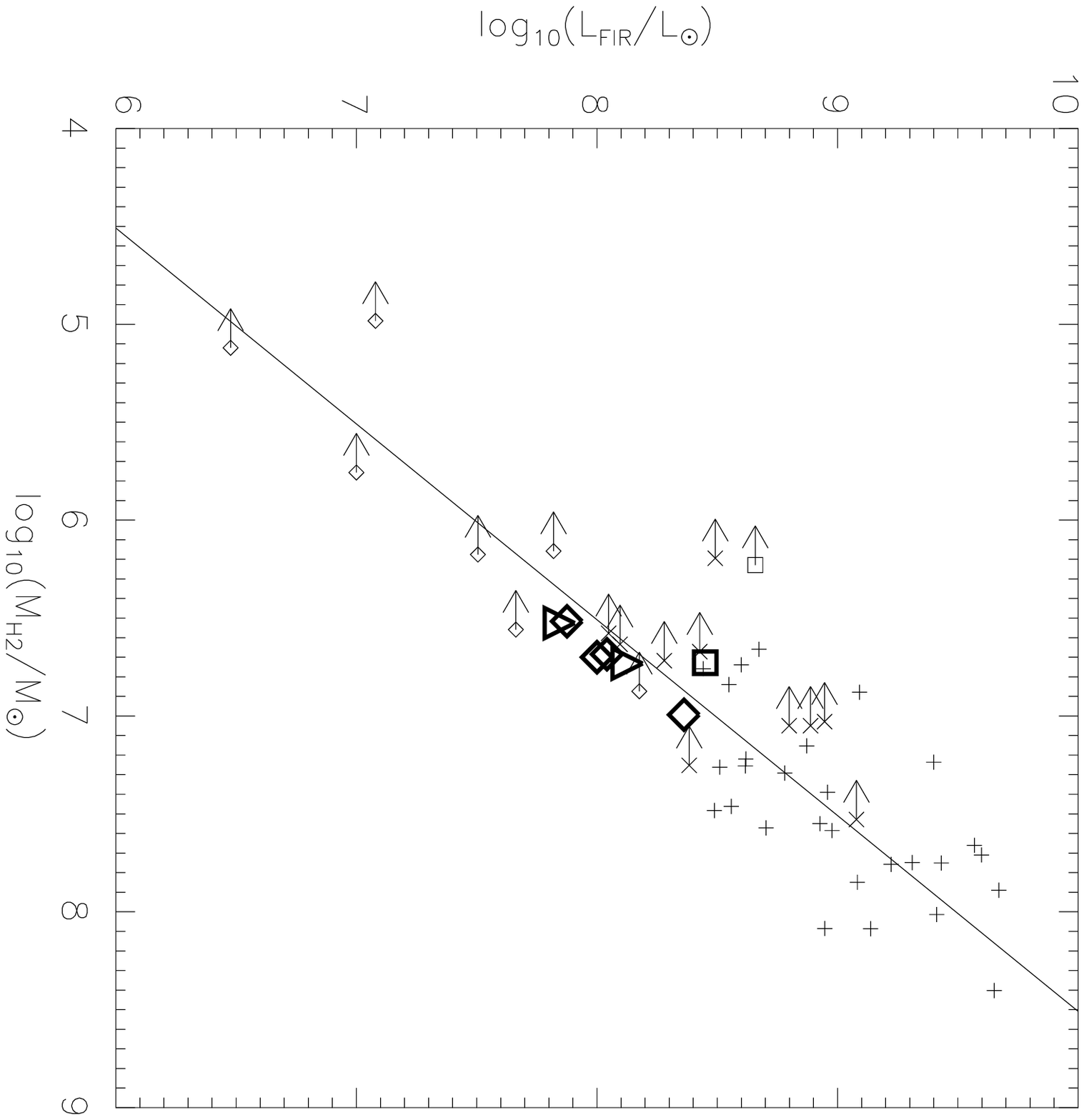}{7.0in}{90.0}{400}{500}{40}{350}
\caption{Logarithm of the  nuclear molecular hydrogen mass versus logarithm of
the far-infrared luminosity, both in solar units, for the
combined LSB spiral sample and the extreme late-type spiral sample of 
Figure~\ref{fig:atomic}. The solid line is not a fit to
the data, but shows the
approximate linear correlation $L_{FIR}/M_{\rm H_{2}}\approx31$ ($L_{\odot}/M_{\odot}$)
derived between the nuclear H$_{2}$ mass
and the total $L_{FIR}$ value by BLS03 for a sample of high surface brightness
spirals (see Section~\ref{FIR}).}
\label{fig:FIR}
\end{figure}

\begin{figure}
\vspace{-16.0cm}
\plotfiddle{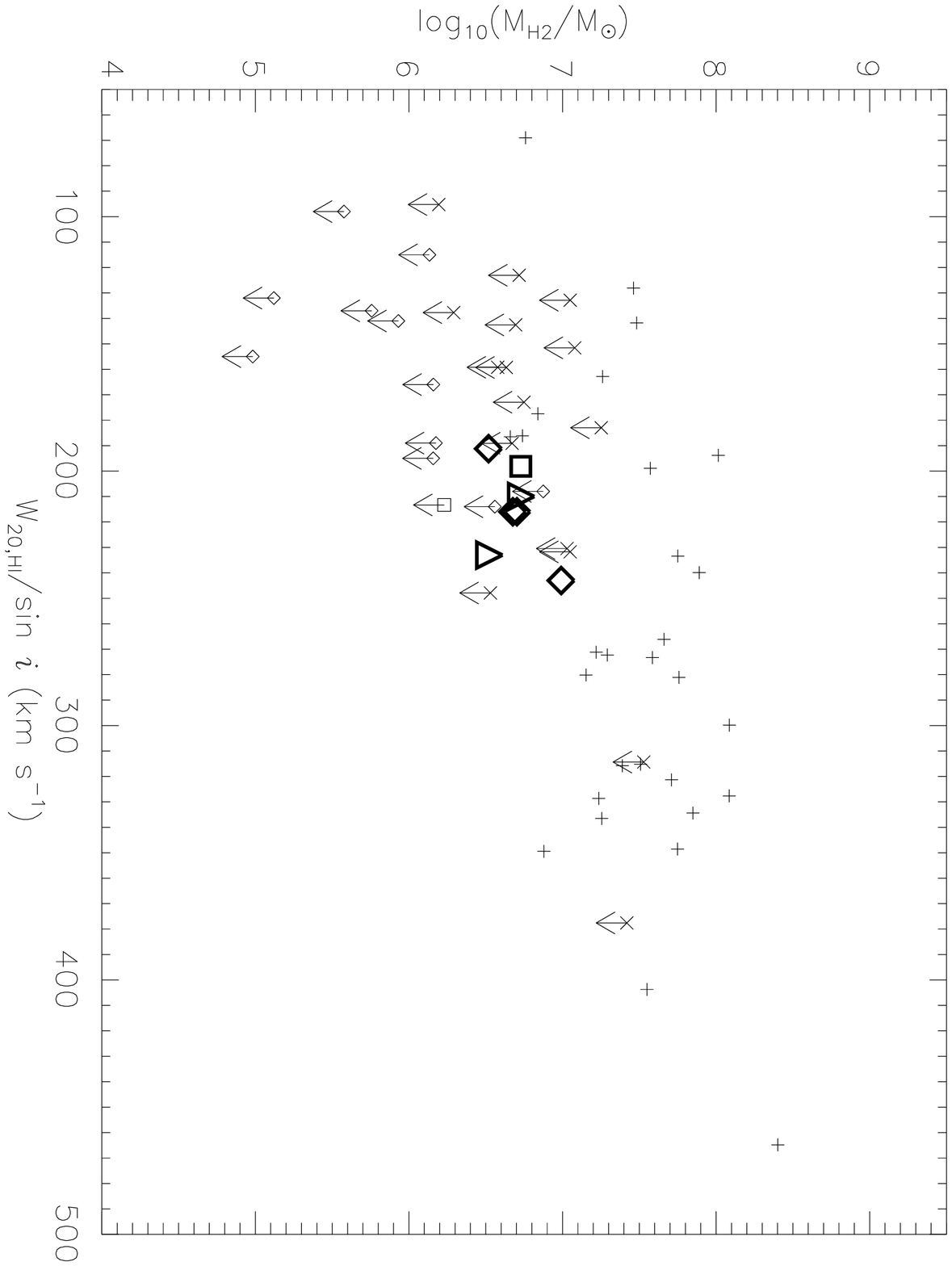}{7.0in}{90.0}{300}{400}{20}{350}
\caption{Inclination-corrected \HI\ linewidth (in \kms) versus 
logarithm of the
nuclear region molecular hydrogen mass, in solar masses,
for the
combined LSB spiral sample and the extreme late-type spiral sample of 
Figure~\ref{fig:atomic}. }
\label{fig:w20}
\end{figure}

\begin{figure}
\vspace{-16.0cm}
\plotfiddle{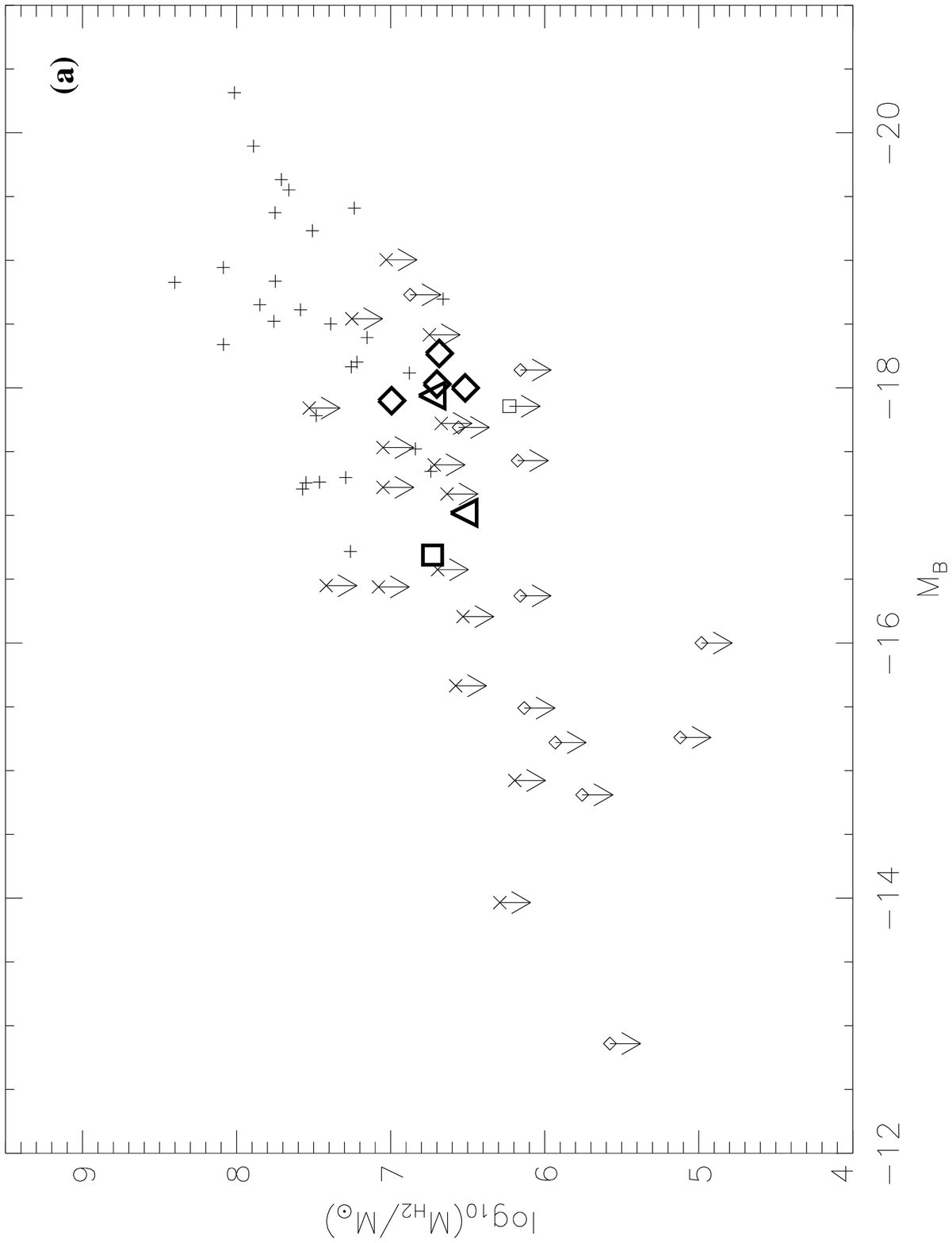}{8.0in}{-90.0}{350}{450}{-20}{350}
\caption{Panel $a$: blue absolute magnitude versus 
logarithm of the nuclear molecular hydrogen mass (in solar
masses) for the
combined LSB spiral sample and the extreme late-type spiral sample of 
Figure~\ref{fig:atomic}.}
\end{figure}

\addtocounter{figure}{-1}
\begin{figure}
\vspace{-16.0cm}
\plotfiddle{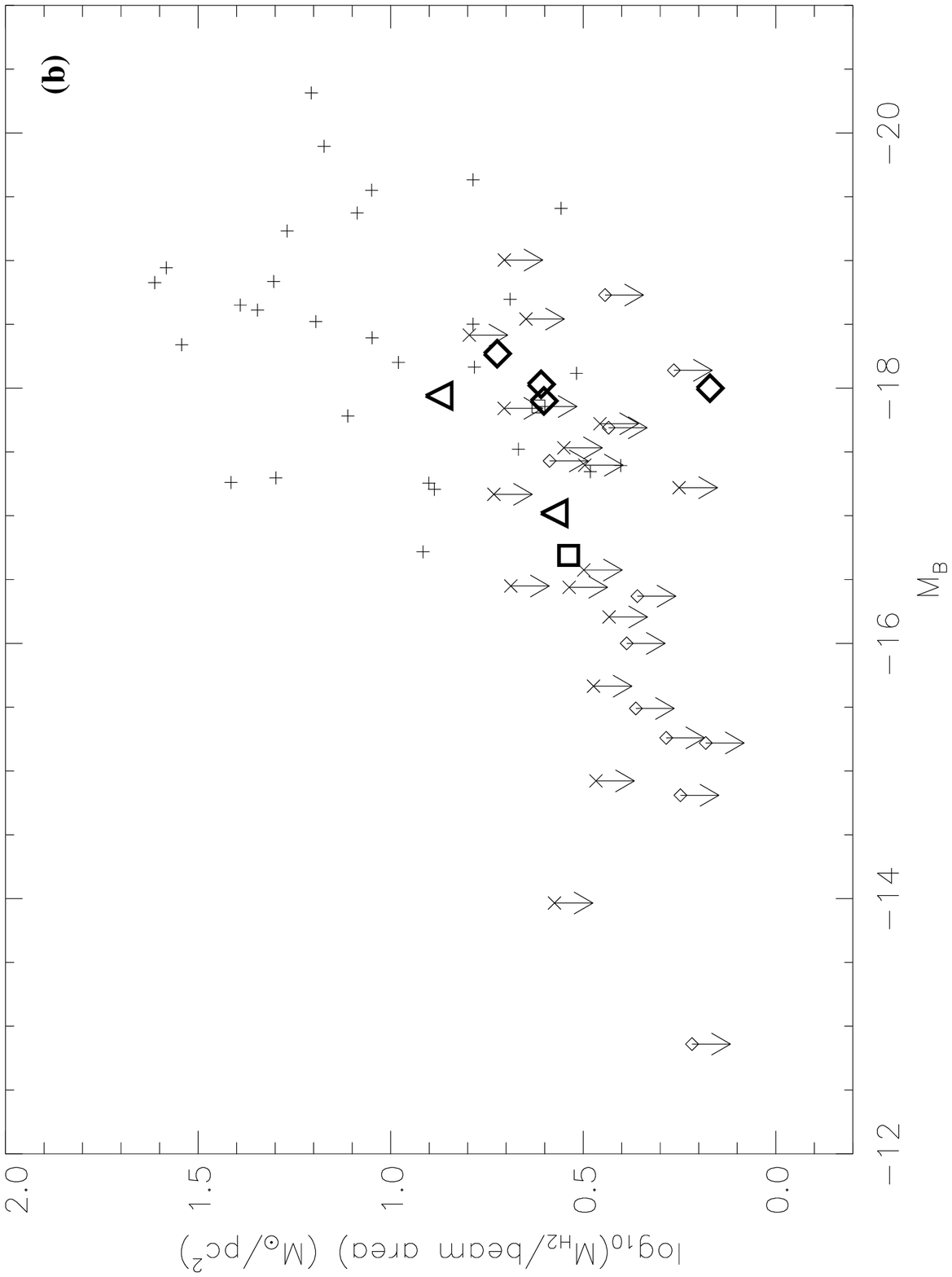}{8.0in}{-90.0}{350}{450}{-20}{350}
\caption{cont. Panel~$b$: blue absolute magnitude versus
logarithm of the central H$_{2}$ surface density (in solar masses per
square parsec).}
\label{fig:metalstest}
\end{figure}

\begin{figure}
\vspace{-15.0cm}
\plotfiddle{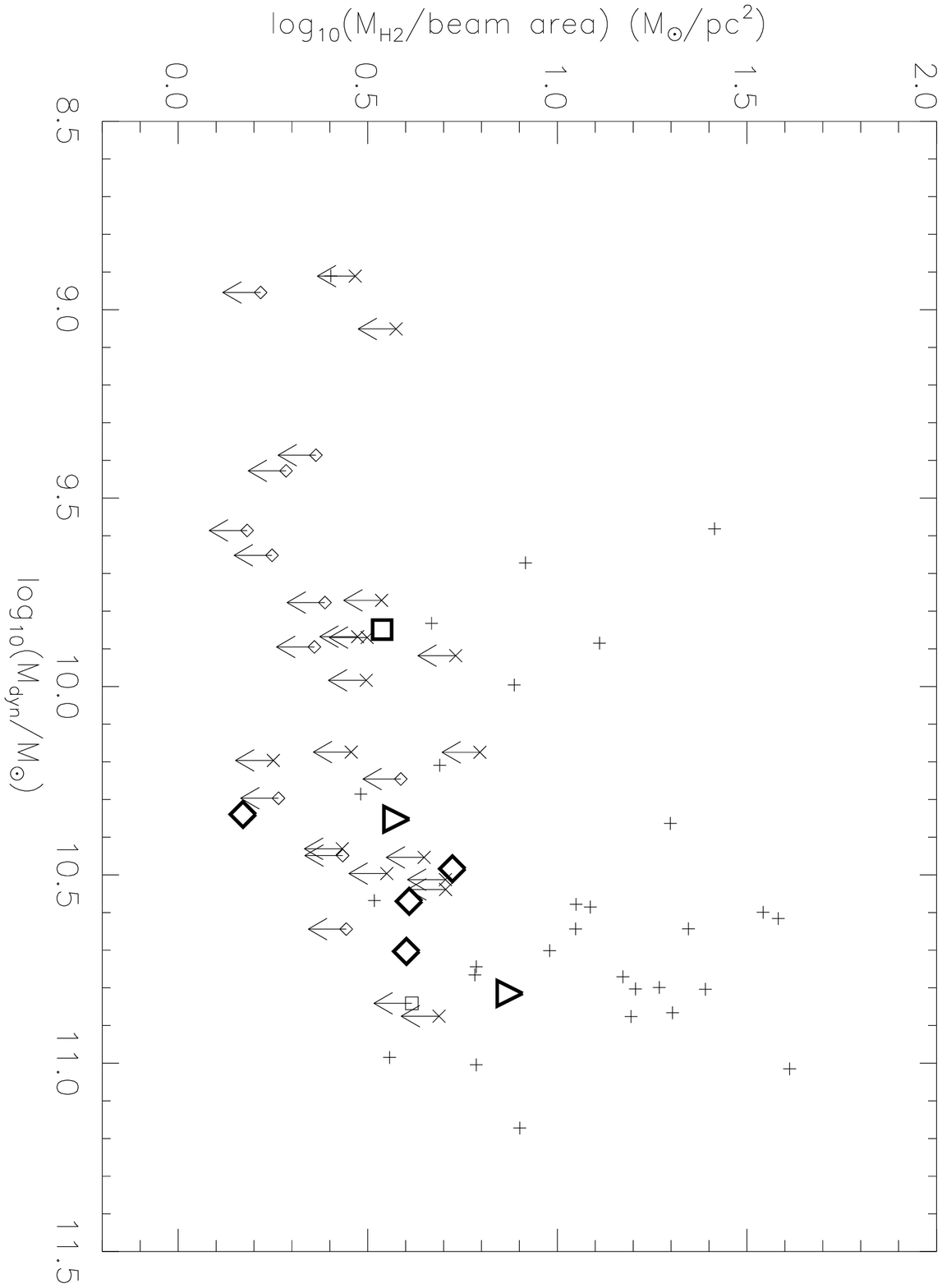}{7.0in}{90.0}{300}{400}{30}{350}
\caption{Logarithm of the dynamical mass (in solar
units) versus logarithm of the central H$_{2}$ surface density 
(in solar masses per
square parsec) for the
combined LSB spiral sample and the extreme late-type spiral sample of 
Figure~\ref{fig:atomic}. }
\label{fig:mdyn}
\end{figure}

\end{document}